\documentclass[twocolumn,superscriptaddress,prb]{revtex4}
\usepackage{amsmath}
\usepackage{amssymb}
\usepackage{amsfonts}
\usepackage{graphicx,bm}
\usepackage{hyperref}

\input{tcilatex}

\begin{document}

\title{Evidence for coherent quantum phase-slips across a Josephson junction
array}
\author{Vladimir E. Manucharyan}
\affiliation{Departments of Physics and Applied Physics, Yale University, New Haven, CT
06520, USA.}
\affiliation{Society of Fellows, Harvard University, Cambridge, MA 02138.}
\author{Nicholas A. Masluk}
\affiliation{Departments of Physics and Applied Physics, Yale University, New Haven, CT
06520, USA.}
\author{Archana Kamal}
\affiliation{Departments of Physics and Applied Physics, Yale University, New Haven, CT
06520, USA.}
\author{Jens Koch}
\affiliation{Departments of Physics and Applied Physics, Yale University, New Haven, CT
06520, USA.}
\affiliation{Department of Physics and Astronomy, Northwestern University, Evanston, IL
60208.}
\author{Leonid I. Glazman}
\affiliation{Departments of Physics and Applied Physics, Yale University, New Haven, CT
06520, USA.}
\author{Michel H. Devoret}
\affiliation{Departments of Physics and Applied Physics, Yale University, New Haven, CT
06520, USA.}

\begin{abstract}
Superconducting order in a sufficiently narrow and infinitely long wire is
destroyed at zero temperature by quantum fluctuations, which induce $2\pi$
slips of the phase of the order parameter. However, in a finite-length wire
coherent quantum phase-slips would manifest themselves simply as shifts of
energy levels in the excitations spectrum of an electrical circuit
incorporating this wire. The higher the phase-slips probability amplitude,
the larger are the shifts. Phase-slips occurring at different locations
along the wire interfere with each other. Due to the Aharonov-Casher effect,
the resulting full amplitude of a phase-slip depends on the offset charges
surrounding the wire. Slow temporal fluctuations of the offset charges make
the phase-slips amplitudes random functions of time, and therefore turn
energy levels shifts into linewidths. We experimentally observed this effect
on a long Josephson junction array acting as a \textquotedblleft
slippery\textquotedblright\ wire. The slip-induced linewidths, despite being
only of order $100~\mathrm{kHz}$, were resolved from the flux-dependent
dephasing of the fluxonium qubit.
\end{abstract}

\maketitle

%
%

\section{Introduction}

The basic notion of superconductivity as a cooperative phenomenon is the
superconducting order parameter. It is a continuous and complex
single-valued function of coordinates, commonly referred to as the
macroscopic wave function\cite{GinzburgLandau1950}. The electronic
condensate of a superconductor can move without friction, manifesting itself
as a current without Joule dissipation. The density of a non-dissipative
current (supercurrent) is proportional to the phase gradient of the
macroscopic wave function; its amplitude is constant along the wire. At
fixed gradient, phase is a linear function of the coordinate along the wire.
For a long wire, the phase difference between the wire's ends may exceed, by
many orders of magnitude, the basic period $2\pi $. This behavior is
inconsistent with the thermodynamics of a superconductor, as its free energy
is a $2\pi $-periodic function of the phase difference. Thus, the
equilibrium current reaches a maximum at a phase difference of $\pi $ and
then oscillates with a further increase of phase difference.

In a narrow wire (thinner than the coherence length), the adjustment of the
supercurrent to the equilibrium value is achieved by the transient processes
of $2\pi $ phase-slips (PS) of the macroscopic wave function~\cite%
{HalperinPhaseSlipsReview}. As the wire becomes thinner, the phase-slips
occur more frequently. Furthermore, at low temperature and for sufficiently
small wire diameters, quantum phase-slips (QPS) take over the thermally
activated ones, leading to an activationless relaxation of supercurrent~\cite%
{Zaikin97}.

It is important to notice that the conditions of the continuity and
single-valuedness of the macroscopic wave function allow a $2\pi m_{i}$
discontinuity of its phase at any point $x_{i}$ along the wire (here $m_{i}$
are integers). At fixed phase difference between the ends of the wire, an
observable quantity, such as current, depends on $m=\sum_{i}m_{i}$, but is
independent of the specific locations $x_{i}$. This is why $m$ can be used
to label different quantum states of the condensate. A $\pm 2\pi $-phase
slip, occurring at some point along the wire, changes the value of $m$ by $%
\pm 1$. Because the state of the wire is characterized by $m=\sum_{i}m_{i}$
rather than by each $m_{i}$ separately, \textit{any} of these $\pm 2\pi $%
-QPS results in the transition $m\rightarrow m\pm 1$. The QPS processes a
priori do not have to be dissipative~\cite{Buchler2004, MooijNazarov2006}.
In the absence of dissipation, QPS happening at different points along the
wire interfere with each other~\cite{Ivanov, MLG2002}. The resulting
superpositions of QPS depend on the distribution of the electric charge
along the wire due to Aharonov-Casher effect~\cite{AC}. We refer to these
spatially interfereing QPS as \textit{coherent} quantum phase-slips (CQPS).

The thinner the wire, the larger the amplitude of CQPS and quantum
uncertainty of $m$ are. Proliferation of CQPS destroys superconductivity in
ultra-thin long wires in the following sense: the equilibrium maximum
supercurrent (prescribed by the periodic phase dependence of the free
energy) decreases exponentially with the increasing length of the wire~\cite%
{MLG2002}, rather than being inversely proportional to it. In general, CQPS
are considered to be the precursors of the superconductor-insulator quantum
phase transition~\cite{FazioVanDerZant2001}. Moreover, in Josephson networks
with special symmetries, CQPS were predicted to give rise to the
topologically non-trivial quantum collective order~\cite%
{DoucotVidal2002,IoffeFeigelman2002}. On a practical side, challenging, but
ultimately achievable control of external dissipation promises realization
of a fundamental current standard by Bloch oscillations induced by CQPS~\cite%
{AverinZorinLikharev, Haviland2000, Buchler2004, MooijNazarov2006}.

In spite of their conceptual and practical importance, the CQPS processes
have remained so far rather elusive. Several experiments have focused on the
resistance of a current-biased superconducting wire, which therefore
dissipates an energy $I\Phi _{0}$ per phase-slip event ($I$ being the bias
current, $\Phi _{0}=h/2e$ the flux quantum)\cite{BezryadinLauTinkham2000,
ArutyunovGolubevZaikin, Rogachev2009}. Such experiments, while providing
evidence for quantum-mechanical effects, cannot reveal the coherent aspects
of quantum phase-slips and are likely to suffer from uncontrolled
dissipation in the electromagnetic environment of the biasing circuitry.
Phase-biased chains of a few Josephson junctions exhibit suppression of
maximum supercurrent consistent with the CQPS theory\cite%
{Gladchenko2009,Pop2010}, but have so far not displayed the expected
spectrum of excitations associated with the quantum dynamics of phase-slips.
Proposals for non-dissipative experiments with nanowires undergoing CQPS in
the flux-qubit setup~\cite{MooijHarmans, MooijTHY} require a narrow range of
quantum phase-slip amplitudes corresponding to transition frequencies of the
order of a few~$\mathrm{GHz}$. Such experimental strategy appears feasible,
but it may be complicated by the fact that phase-slip amplitudes of
realistic wires vary by many orders of magnitude in view of their predicted
exponential sensitivity to the wire transverse dimensions~\cite{Zaikin97}.

In this paper we describe an evidence for the coherent quantum phase-slips
in a long array of $N=43$ Josephson junctions, which emulates a nanowire.
Our experiment on CQPS features two important distinctions. First, for a
faithful emulation of a nanowire, two parameters of the array are crucial:
it must be long (comprising many junctions) and the probability amplitude of
a phase-slip in a single junction must be small. These conditions minimize
the effect of the discrete nature of the lumped circuit, and make the
dynamics of condensate phase in the array similar to its counterpart in a
wire. Second, to detect the presence of CQPS in our array, we take advantage
of the sensitivity of the fluxonium qubit\cite{Manucharyan2009} transitions
to the phase-slips in its array inductance, which we treat as a slippery
superconducting wire.

This paper is organized as follows. In section~\ref{Section2} of this paper
we describe, both qualitatively and quantitatively, how the spectrum of
fluxonium qubit is modified by the presence of CQPS in the qubit inductance.
In section \ref{Section3} we describe the experimental evidence of CQPS in
the qubit inductance by comparing to the theory both frequency-domain and
time-domain measurements of the fluxonium transitions. Appendix A provides
details of our experimental techniques, Appendix B supports our theory and
data analysis.

\section{Dispersive effect of quantum phase-slips on a superconducting
artificial atom}

\label{Section2}

\begin{figure*}[tbp]
\resizebox{\linewidth}{!}{\includegraphics{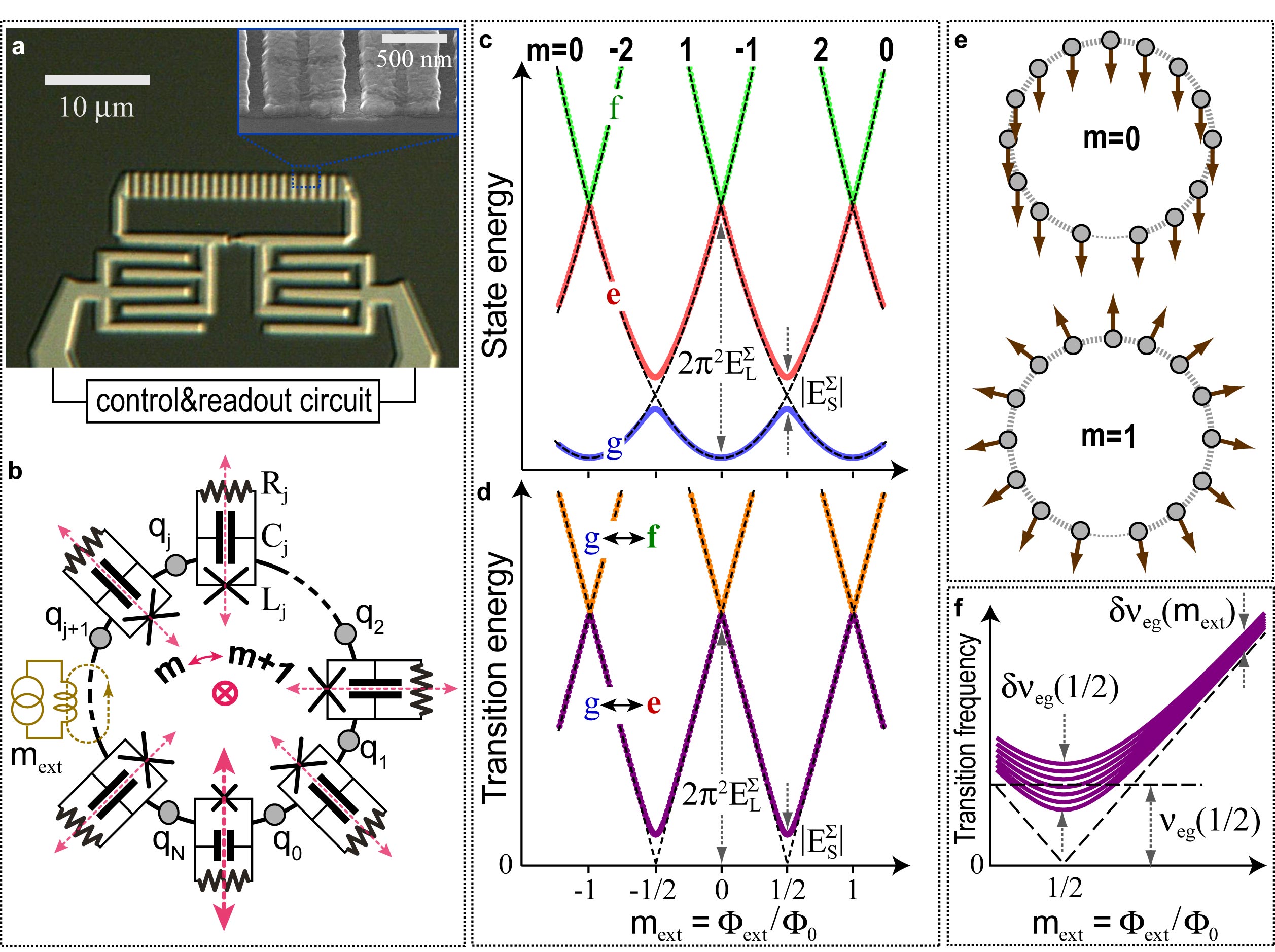}}
\caption{(a) Optical image of the fluxonium device, bright color for Al,
dark color for Si substrate. Two finger capacitors couple the small junction
of the loop to the dispersive microwave reflectometry setup. Inset: Scanning
electron micrograph of an array section. (b) Effective electrical circuit of
the device with minimal elements for quantum phase-slip description. Each
junction consists of a non-linear Josephson inductance $L_{j}$, a
capacitance $C_{j}$, and an effective frequency-dependent resistance $R_{j}$
representing possible intrinsic losses; every island carries offset charge $%
q_{j}$ induced by charged impurities in the oxide or unpaired
quasiparticles. Pink arrows indicate possible semiclassical phase-slip
trajectories. Charge-dependent energy shifts of artificial atom levels
correspond to closed trajectories crossing several junctions (not
represented on the figure). (c) Low-energy spectrum of the Josephson ring.
Dashed parabolas represent energy of loop Josephson inductance threaded by
an integer number $m$ of flux quanta and tuned with external flux $m_{%
\mathrm{ext}}$. (d) Two lowest transitions from the ground state. (e) Sketch
of spatial distribution of superconducting phases of array islands,
corresponding to the states with $m=0$ (no phase-slips) and $m=1$ (one
phase-slip) across the entire loop at $m_{\mathrm{ext}}=0$. Phase of each
island is represented by the angle between the corresponding arrow and the
vertical. The value of $m$\ coinsides with the total phase winding along the
loop devided by $2\protect\pi $. (f) Schematic broadening of the\textbf{\ }%
ground to first excited state transition $g\rightarrow e$ due to a spread of
CQPS amplitude across the array. Origin of the spread is the combination of
Aharonov-Casher effect and time-dependence of offset charges $q_{j}$ of
array islands. Average transition frequency at $m_{\mathrm{ext}}=1/2$ is
dominated by the phase-slip across the weak junction.}
\label{Figure1}
\end{figure*}

Fluxonium qubit (Fig.~\ref{Figure1}a, Fig.~\ref{FIG-DevicePhotograph}) may
be viewed as one junction shunted by the kinetic inductance of a series
array of $N\gg 1$ larger junctions~\cite{Manucharyan2009}. Alternatively, we
may view it as a loop of $N+1$ junctions connected in series, with one
\textquotedblleft black-sheep\textquotedblright\ weaker junction (Fig.~\ref%
{Figure1}b). The quantum state of the device is monitored by coupling the
weak junction capacitively to an electromagnetic resonator, which provides a
non-dissipative microwave readout\cite{Schuster2005, Wallraff04}.

The loop array can first be understood qualitatively by the following
semiclassical approach. Quantum fluctuations of the phase across a selected $%
j$-th Josephson junction are controlled by the ratio of its Josephson energy
($E_{J_{j}}=(\Phi _{0}/2\pi )^{2}/L_{j}$, $L_{j}$ being junction inductance)
and charging energy ($E_{C_{j}}=e^{2}/2C_{j}$, $C_{j}$ being junction
capacitance). We neglect the capacitances of the array islands to ground
(any large size metal object in the vicinity of the device). This is made
possible by a tight packing of array junctions (Fig.~\ref{Figure1}a-inset),
as justified in our previous work~\cite{Manucharyan2009}.

For large $E_{J_{j}}/E_{C_{j}}$, phase fluctuations are small.
Correspondingly, quantum phase-slips crossing a ring made of such junctions
occur rarely. This allows one to reduce the many-body Hamiltonian describing
the quantum dynamics of phases in the ring to a simplified, low-energy
Hamiltonian~\cite{MLG2002}. In the basis of states labeled by the number $m$
of phase-slips which crossed the ring the Hamiltonian is given by (model A) 
\begin{align}
H& =\frac{1}{2}E_{L}^{\Sigma }(2\pi )^{2}\sum\limits_{m}(m-m_{\mathrm{ext}%
})^{2}\left\vert m\right\rangle \left\langle m\right\vert  \notag \\
& +\sum\limits_{m}\left[ \frac{1}{2}E_{S}^{\Sigma }\left\vert m\right\rangle
\left\langle m+1\right\vert +\mathrm{h.c.}\right] \,.
\label{BasicPhaseSlipsHamiltonian}
\end{align}%
Here $E_{L}^{\Sigma }=\left( \Phi _{0}/2\pi \right) ^{2}/L_{\Sigma }$
corresponds to the inductive energy of the current in the loop of total
inductance $L_{\Sigma }=\sum L_{j}$; the hybridization matrix element $%
E_{S}^{\Sigma }$ is proportional to the CQPS probability amplitude in the
loop, and we assume $|E_{S}^{\Sigma }|\ll $ $E_{L}^{\Sigma }$, so that one
can neglect non-nearest-neighbor terms (like $m\leftrightarrow m+2$); an
externally applied magnetic field offsets the total flux in the loop by an
amount $\Phi _{\mathrm{ext}}=$ $m_{\mathrm{ext}}\Phi _{0}$.

In the absence of CQPS, i.e. $E_{S}^{\Sigma }=0$, the energy levels of the
ring depend quadratically on $m_{\mathrm{ext}}$ for a given $m$ (Fig.~\ref%
{Figure1}c), and present two-fold degeneracy at half-integer $m_{\mathrm{ext}%
}$. The energy difference between the ground $g$ and first excited $e$
states varies in a zig-zag manner as a function of $m_{\mathrm{ext}}$. CQPS
removes the degeneracy, which leads to the rounding of the zig-zag corners
at half-integer $m_{\mathrm{ext}}$. Moreover, the presence of CQPS allows $%
e\leftrightarrow g$ coherent transitions under external radiation at any $m_{%
\mathrm{ext}}$.

Interestingly, the collective nature of the strongly-coupled junctions in
this model hides in the specific form of the matrix element $E_{S}^{\Sigma }$%
. Namely, 
\begin{equation}
E_{S}^{\Sigma }\{Q\}=\sum\limits_{j=0}^{N}E_{S_{j}}\mathrm{e}^{i2\pi
Q_{j}/2e}\,,  \label{PhaseSlipsInterference}
\end{equation}%
where $E_{S_{j}}$ is the \textquotedblleft microscopic\textquotedblright\
contribution of an individual quantum phase-slip along the $j$-th junction%
\cite{ESnote} and $Q_{j}$ is the total charge on the islands between the $0$%
-th and $j$-th junction, i.e $Q_{j}=\sum\limits_{i=0}^{j}q_{i}$, with $q_{j}$
being the charge on $j$-th island (Fig.~\ref{Figure1}b). As long as $N\gg 1$%
, one can view each quantum phase-slip event as tunneling of a fictitious
particle carrying flux $\Phi _{0}$ from the inside of the loop to the
outside (and vice-versa). Tunneling occurs via a superposition of multiple
paths crossing different junctions and therefore encircle island charges.
Tunneling through each individual junction $j$ then contributes to the total
CQPS amplitude Eq.~(\ref{PhaseSlipsInterference}) with the corresponding
Aharonov-Casher geometric phase~\cite{AC,Elion,AverinFriedman,Ivanov,MLG2002}
$2\pi Q_{j}/2e$.

In real superconducting circuits, however, non-equilibrium charged
impurities and quasiparticles cause the offset charges to fluctuate in time
with an amplitude comparable to $e$, often on sub-$\mathrm{ms}$ time scales~%
\cite{Aumentado2004}, much shorter than the averaging time of a typical
qubit experiment. In the case of a homogeneous array $(E_{S_{i}}\simeq
E_{S_{j}}$ $\forall i,j)$, fluctuations of $E_{S}^{\Sigma }$ and, as a
result, fluctuations of the transition matrix element $\left\langle
g|m|e\right\rangle $, which couples the qubit to the external radiation, are
comparable to their respective means (see Appendix~\ref{ACbroadening}). This
makes slow spectroscopic measurement at $m_{\mathrm{ext}}=1/2$, and
time-domain coherence measurements at any $m_{\mathrm{ext}}$ technically
challenging. We circumvent this obstacle by introducing an inhomogeneity in
the array in the form of one ($j=0$, for definiteness) weak
\textquotedblleft black-sheep\textquotedblright\ junction. Since the phase
across the weaker junction fluctuates more strongly, the corresponding
phase-slip amplitude $E_{S_{0}}$ at $j=0$ largely exceeds that for all other
(array) junctions, $E_{S_{0}}\gg E_{SA}\equiv \overline{E_{S_{j\neq 0}}}$,
where the averaging is taken over the junction index. Now, the
Aharonov-Casher interference contrast is reduced from unity to a much
smaller value of order $\sqrt{N/2}E_{SA}/E_{S_{0}}$, where the $\sqrt{N}$
factor comes from averaging over random charges. Effectively, we split the
roles of phase-slips in different junctions: while phase-slips across the
black-sheep mix states of the loop with different $m$ and therefore shape
the qubit transition spectrum (Fig.~\ref{Figure1}b-c), the CQPS in the array
junctions, combined with the fluctuating island charges, induce an
inhomogeneous broadening $\delta \nu _{eg}$ to the qubit\textbf{\ }$%
g\leftrightarrow e$ transition. This linewidth is maximal at $|m_{\mathrm{ext%
}}|=1/2$, where it is given by $\delta \nu _{eg}(1/2)=\sqrt{N/2}E_{SA}/h$,
and diminishes away from that spot according to the following expression 
\begin{equation}
\delta \nu _{eg}(m_{\mathrm{ext}})=\delta \nu _{eg}(1/2)\frac{\nu _{eg}(1/2)%
}{\nu _{eg}(m_{\mathrm{ext}})}  \label{PhaseSlipsLinewidthSimple}
\end{equation}%
(Fig.~\ref{Figure1}f). The factor $\sqrt{N/2}$ comes from averaging $%
|E_{S}^{\Sigma }|$ over the random charges, assuming $E_{S_{0}}\gg E_{SA}$.

CQPS thus remarkably turn the flux-noise sweet spot into charge noise
anti-sweet spot. We have now arrived at the main idea behind our experiment:
by measuring the variation of the dephasing time $T_{2}^{\ast }$ of the $%
g\leftrightarrow e$ transition of the fluxonium circuit as a function of
external flux $m_{\mathrm{ext}}$, we resolve the effect of CQPS in the
Josephson junction array, provided that the transition intrinsic linewidth
is smaller than the characteristic frequency $\sqrt{N/2}E_{SA}/h$ of the
CQPS. Evidence for CQPSs consists of: (i) the presence of the dephasing
anti-sweet spot at $|m_{\mathrm{ext}}|=1/2$, (ii) excellent quantitative
agreement of the observed dephasing time as a function of $m_{\mathrm{ext}}$
with the described below parameter-free prediction Eq.~(\ref{A-Clinewidth}),
and (iii) confirmation of the inhomogeneous nature of the dephasing in echo
measurements. The latter also allow us to extract limits for the time scale
of charge re-arrangements.

We introduce now a quantitative model of quantum phase-slips in the
fluxonium circuit (model B). Unlike Eq.~(\ref{BasicPhaseSlipsHamiltonian}),
now we need to allow strong phase fluctuations in one of the junctions,
while phase-slips in all others are still rare. Therefore, we use a mixed
representation: the black-sheep junction is described by a continuous
fluctuating phase $\varphi $; the dynamics of the remaining array of large
junctions is treated by the same tight-binding type model as Eq.~(\ref%
{BasicPhaseSlipsHamiltonian}) in the space of the number $\tilde{m}$ of CQPS
across the array of large junctions only. The variables $\varphi $ and $%
\tilde{m}$ are coupled because phase-slips in the array of large junctions
create a phase bias on the black sheep. Thus, the Hamiltonian of model B is 
\begin{align}
H& =H_{F}(\varphi ,~m_{\mathrm{ext}}-\tilde{m})  \notag \\
& +\sum\limits_{\tilde{m}}\left[ \frac{1}{2}E_{S}\left\vert \tilde{m}%
\right\rangle \left\langle \tilde{m}+1\right\vert +\mathrm{h.c.}\right] \,,
\label{MainHamiltonianPart1}
\end{align}%
where 
\begin{align}
H_{F}(\varphi ,m_{\mathrm{ext}})& =-4E_{C}\frac{\mathrm{\partial }^{2}}{%
\mathrm{\partial }\varphi ^{2}}-E_{J}\cos \varphi   \notag \\
& +\frac{1}{2}E_{L}(2\pi )^{2}(\varphi /2\pi -m_{\mathrm{ext}})^{2}\,,
\label{MainHamiltonianPart2}
\end{align}%
$E_{J}=E_{J_{0}}$, and $E_{C}=E_{C_{0}}$. The CQPS amplitude $%
E_{S}\{Q\}=E_{S}^{\Sigma }\{Q\}-E_{S_{0}}$ is different from that in Eq.~(%
\ref{PhaseSlipsInterference}) and accounts for phase-slips through every
junction except the black-sheep one (we have set $Q_{0}=0$ without the loss
of generality). Similarly, the new inductive energy $E_{L}$ excludes the
contribution of the black-sheep junction inductance from the total inductive
energy of the loop; $1/E_{L}=$ $1/E_{L}^{\Sigma }-1/E_{J_{0}}$. For $E_{S}=0$%
, expressions (\ref{MainHamiltonianPart1}) and (\ref{MainHamiltonianPart2})
define the Hamiltonian of the fluxonium qubit\cite{Manucharyan2009}. The
second term in~Eq~(\ref{MainHamiltonianPart1}) incorporates the CQPS in the
fluxonium inductance, and $\tilde{m}$ now counts their number. We are now in
a position to evaluate accurately the flux dependence of the Aharonov-Casher
linewidth of any $\alpha \leftrightarrow \beta $ fluxonium transition. First
order perturbation theory in $|E_{S}|$ results in a remarkably simple
expression (see Appendix~\ref{ACbroadening} for details of derivation): 
\begin{equation}
\delta \nu _{\alpha \beta }(m_{\mathrm{ext}})=\frac{E_{SA}\sqrt{N/2}}{h}%
\left\vert F_{\alpha \beta }(m_{\mathrm{ext}})\right\vert \,,
\label{A-Clinewidth}
\end{equation}%
where 
\begin{align}
F_{\alpha \beta }(m_{\mathrm{ext}})& =\int\limits_{-\infty }^{\infty }%
\mathrm{d}\varphi \Psi _{\alpha }(\varphi )\Psi _{\alpha }(\varphi -2\pi ) 
\notag \\
& -\int\limits_{-\infty }^{\infty }\mathrm{d}\varphi \Psi _{\beta }(\varphi
)\Psi _{\beta }(\varphi -2\pi )  \label{A-Clinewidth2}
\end{align}%
with $\Psi _{\alpha }(\varphi )$ being the eigenfunction of the $\alpha $-th
energy state of the fluxonium Hamiltonian~(\ref{MainHamiltonianPart2}); it
can be readily computed numerically.

The dependence of $\delta \nu _{\alpha \beta }$ on the external flux $m_{%
\mathrm{ext}}$ comes from the presence of $m_{\mathrm{ext}}$ in Eq.~(\ref%
{MainHamiltonianPart2}) determining the wave functions $\Psi _{\alpha ,\beta
}(\varphi )$ which enter Eq.~(\ref{A-Clinewidth2}). For small amplitude of
CQPS and for $|m_{\mathrm{ext}}|$ close to $1/2$, one recovers the relation~(%
\ref{PhaseSlipsLinewidthSimple}) from Eq.~(\ref{A-Clinewidth}). The overlap
function $F_{\alpha \beta }$ also appears in a previous work\cite%
{HriscuNazarov} and can be understood from the fact that a phase-slip
through the array inductance must shift the center of gravity of $\Psi
_{\alpha }(\varphi )$ by $2\pi $. The external flux dependence of the
Aharonov-Casher linewidth thus encodes the overlap of the $2\pi $-shifted
fluxonium wave functions (Fig.~\ref{FIG-WafeFunctionOverlap}). The flux
dependence of the linewidth$~$(\ref{A-Clinewidth}) of our model B
generalizes Eq.~(\ref{PhaseSlipsLinewidthSimple}) of the intuitive model A
to the case of arbitrary black-sheep junction parameters.

Finally, since the finite lifetime of fluxonium excited states limit our
experiment, and given the large number $N$ of array junctions, a natural
question arises: does the dissipation in the array scale up with $N$? Every
junction of the array suffers from intrinsic, generally unknown dissipation
sources, all lumped into a frequency-dependent resistance $R_{j}$ shunting
the $j$-the junction (Fig.~\ref{Figure1}b). Fortunately (and
counterintuitively) the dissipation in the black-sheep junction solely
dominates the intrinsic relaxation of the phase-slips spectrum with
contribution from array junctions being suppressed as $1/N$. Indeed, the
dominant phase-slip process generates a voltage pulse across the
black-sheep; each array junction receives only a $1/N$ portion of that
voltage, and the total energy dissipated in the array resistors is only $1/N$
of that dissipated in the black-sheep junction. We thus conclude that,
remarkably, the apparent multi-junction complexity of the fluxonium circuit
does not a priori penalize it with enhanced energy relaxation.

\section{Experiments}

\label{Section3}

In this section we present measurements of the transition frequencies of a
fluxonium qubit obtained using both frequency-domain and time-domain
techniques, and analyze our data using the theory developed in section~\ref%
{Section2}. We read out the qubit state with the help of a microwave
resonator, the technique known as circuit QED~\cite{Schuster2005, Wallraff04}%
. Details of this technique are described in the Appendix, Sections~\ref%
{APP-Reflectometry},~\ref{APP-MeasurementSchematic}, and~\ref%
{APP-DispersiveShifts}. Capacitive coupling of the fluxonium circuit to the
readout resonator results in the shift of the frequency of this resonator
depending on the quantum state of fluxonium according to Eq.~(\ref%
{DispersiveShift}). Essentially, the qubit state is mapped onto a particular
value of the shift of the resonance frequency of the readout resonator. This
frequency shift is in turn detected by measuring the phase (Eq.~(\ref%
{DispersivePhaseShift})) of the reflection amplitude (Eq.~(\ref%
{ReflectionAmplitude})) for a microwave signal, scattered off the resonator.

\begin{figure*}[tbp]
\resizebox{\linewidth}{!}{\includegraphics{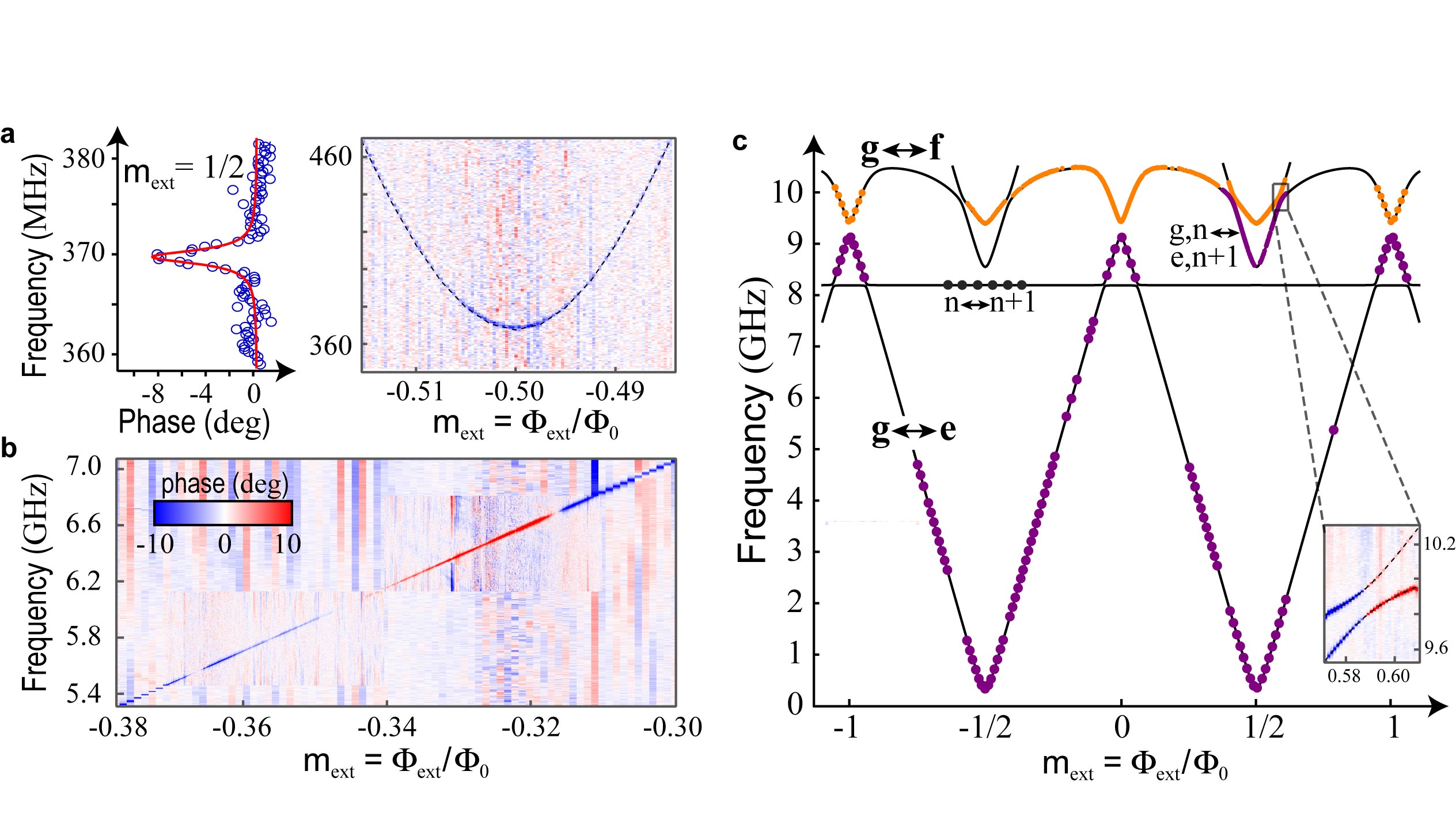}}
\caption{(a) Left: Transition between $g\leftrightarrow e$ fluxonium ground
and first excited states at $|m_{\mathrm{ext}}|=1/2$ measured by sweeping
spectroscopy (Y-axis) frequency and measuring phase response of the $8.2$~$%
\mathrm{GHz}$ microwave cavity coupled to the qubit (X-axis). Right: Raw
spectroscopy data consisting of traces like that on the left but with Y
color-coded, as a function of external flux, in the vicinity of $|m_{\mathrm{%
ext}}|=1/2$; (b) Large scale spectroscopy data as a function of flux and
frequency. (c) Extracted resonance locations vs spectroscopy frequency and
applied flux. Solid line represent the fit to numerically diagonalized
Hamiltonian neglecting quantum phase-slips in the array but taking into
account interaction with readout cavity mode. Horizontal line represents
cavity mode. Inset shows strong, $160~\mathrm{MHz}$ anticrossing of the
cavity-assisted blue sideband of the qubit lowest transition with the second
lowest qubit transition, on which the readout mechanism is based.}
\label{Figure2}
\end{figure*}

\textit{Spectroscopy of fluxonium.} Spectroscopy data in the range of qubit
transition frequencies from $300~\mathrm{MHz}$ to $12$~$\mathrm{GHz}$ and
for the full span of external flux bias reveals the spectrum of the
transitions between the $3$ lowest energy levels (Fig.~\ref{Figure2}). The
phase-slip frequency of the black-sheep junction, $\nu _{eg}(m_{\mathrm{ext}%
}=1/2)=|E_{S_{0}}|/h=369~\mathrm{MHz}$ corresponds to the center frequency
of the line observed at $m_{\mathrm{ext}}=1/2$ (Fig.~\ref{Figure2}a-left).
By varying $m_{\mathrm{ext}}$ around that point and plotting spectroscopy
traces on a 2D color plot (Fig.~\ref{Figure2}a-right) we observe the
anti-crossing of the states of the loop with the phase-slips number $m$\
differing by a unity. Apart from the neighborhood of $m_{\mathrm{ext}}=1/2$
and $m_{\mathrm{ext}}=0$, the transition frequency depends linearly on the
applied flux with a slope given by $E_{L}$, up to corrections of order $%
E_{L}/E_{J_{0}}\simeq 0.06$ (Fig.~\ref{Figure2}b). In $m_{\mathrm{ext}}-\nu $
plane (Fig.~\ref{Figure2}c), one recognizes the anticipated zig-zag shape
(Fig.~\ref{Figure1}d) of the lowest $g\leftrightarrow e$ transition (Fig.~%
\ref{Figure2}c).

\begin{figure*}[tbp]
\resizebox{\linewidth}{!}{\includegraphics{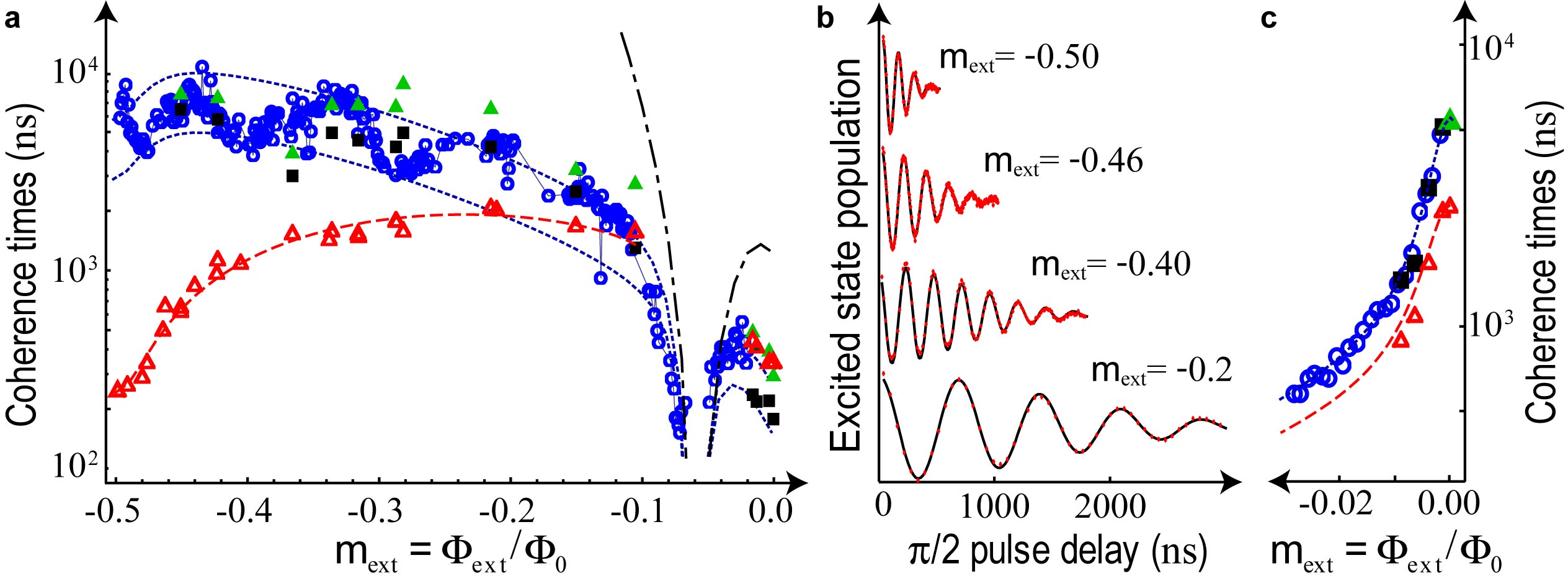}}
\caption{(a) Summary of the coherence times measurements of the $%
g\leftrightarrow e$ transition. Blue open circles denote energy decay times (%
$T_{1}$), red open triangles - decay time of the Ramsey fringes ($%
T_{2}^{\ast }$) and closed green triangles the decay of the $\protect\pi $%
-pulse echo experiment ($T_{2}$). Black squares are the $T_{1}$ data
measured simultaneously with the echo data. The dash-dotted line represent
the calculated relaxation time into the readout cavity (Purcell effect),
which clearly is irrelevant for the most of the data. The two dotted lines
are theoretical predictions of relaxation, for the two absolute values of
effective resistance discussed in the text. Relaxation times $T_{1}$'s
present temporal variations that are not repeatable (differences between
blue open and solid circles), while the overall dependence with applied flux
is. Dashed line corresponds to the theory of the Aharonov-Casher effect
induced dephasing, with the overlap function calculated numerically using
circuit parameters extracted by spectroscopy. (b) Actual Ramsey fringes for
several values of flux bias, with matched scale of the Y-axes for clarity.
(c) Coherence time of the $g\leftrightarrow f$ transition. Notations match
those of (a), except that all relaxation times were multiplied by 2 for easy
comparison with dephasing times. Theory lines obtained using same parameters
as those for $g\leftrightarrow e$ transition.}
\label{Figure3}
\end{figure*}

The anticrossing at $m_{\mathrm{ext}}=0$ between transitions $%
g\leftrightarrow f$ and $g\leftrightarrow e$ (Fig.~\ref{Figure2}c) is
associated with the hybridization between states $m\leftrightarrow m+2$. The
size of this anticrossing according to model A would be orders of magnitude
smaller than $\nu _{eg}(m_{\mathrm{ext}}=1/2)$, the latter is associated
with the hybridization $m\leftrightarrow m+1$. However, we find that the two
frequencies are of the same order, because of the proximity of transitions $%
g\leftrightarrow f$ and $g\leftrightarrow e$ at $m_{\mathrm{ext}}=0$ to the
plasma resonance in the black-sheep junction, which model B properly takes
into account.

The $g\leftrightarrow f$ transition anticrosses the cavity-assisted blue
sideband of the $g\leftrightarrow e$ transition, which appears as a copy of
the $g\leftrightarrow e$ transition (Fig.~\ref{Figure2}c and inset). This
anticrossing is an unusual vacuum Rabi resonance between the readout cavity
and the $e\leftrightarrow f$ transition of the qubit, as indicated by the
pole in the dispersive shift expression (\ref{DispersiveShift}) at $\nu
_{ef}=\nu _{0}$. The magnitude of this anticrossing (Fig.~\ref{Figure2}%
c-inset), which exceeds $100~\mathrm{MHz}$ demonstrates the strong coupling
of the qubit to the readout and accounts for the good qubit visibility in
the remarkable $5$-octave transition frequency range. The frequencies of all
transitions are in perfect agreement with a numerical diagonalization of the
hamiltonian of model B (neglecting $E_{S}$). This allows us to extract
accurately the circuit parameters values from spectroscopy data, see Table~(%
\ref{Table1}).

\textit{Measurement of decoherence times of fluxonium.} We now turn to the
time-domain experiments. The coherence of the $g\leftrightarrow e$
transition (Fig.~\ref{Figure2}a) is analyzed using standard time domain
techniques in the full range of flux and frequency $(-0.5\leqslant m_{%
\mathrm{ext}}\leqslant 0$;$~0.369~\mathrm{GHz}\leqslant \nu _{eg}\leqslant
9.1~\mathrm{GHz})$. First, the contribution of the energy relaxation, is
measured by the standard $\pi $-pulse techniques and yields the exponential
decay times $T_{1}$ of the excited state. Theoretical expression for $T_{1}$
is given by Eqs.~(\ref{Reff}), (\ref{T1Theory}), (\ref{T1Purcell}), and (\ref%
{Rintrinsic}). In the narrow $300~\mathrm{MHz}$ vicinity of the cavity
frequency, $T_{1}$ is limited by relaxation into the cavity (Purcell
effect). Subtracting that effect, the overall shape of the $T_{1}(m_{\mathrm{%
ext}})$ data is matched well (within a factor of $2)$, by dissipation across
the black-sheep junction through effective shunting resistance of the form $%
R_{0}(\omega )=A/\omega $. Note that this agreement takes place for $T_{1}$
varying by two order of magnitude and over a $4-$octave frequency span. The
only adjustable parameter is the value of $A=190\pm 60$~$\mathrm{M\Omega }%
\times \mathrm{GHz}$, the extreme values corresponding to top and bottom
dashed lines (Fig.~\ref{Figure3}a). Such $1/\omega $ frequency dependent
resistance usually arises from the coupling to a large ensemble of discrete
energy absorbers\cite{Martinis}, whose microscopic origin is presently
unclear \cite{FaoroIoffe}. Dissipation across the black-sheep may likely
come from dielectric losses in the coupling finger capacitors (Fig.~\ref%
{Figure1}a). Alternatively, our energy relaxation data could be explained by
a lossless black-sheep and dissipation in the larger area junctions of the
array, but with the $A$ factor $N=43$ stronger.

Dephasing times $T_{2}^{\ast }$ of the $g\leftrightarrow e$ transition,
measured from the decay of Ramsey fringes, display pronounced minimum of
about $250$~$\mathrm{ns}$ at the flux sweet-spot and spectacularly rise by
almost an order of magnitude (Fig.~\ref{Figure3}a-b), exceeding $2$~$\mathrm{%
\mu s}$ at $|m_{\mathrm{ext}}|\simeq 0.2$ ($\nu _{ge}=$ $5.5$~$\mathrm{GHz}$%
). Around $|m_{\mathrm{ext}}|=1/2$, the decay of Ramsey fringes is well
fitted (Fig.~\ref{Figure3}b) with a gaussian. This confirms the irrelevance
of the energy relaxation for $|m_{\mathrm{ext}}|>0.2$, while it dominates at 
$m_{\mathrm{ext}}$ close to zero. Coherence times $T_{2}$ obtained with echo
experiments are drastically larger than $T_{2}^{\ast }$, particularly around 
$|m_{\mathrm{ext}}|=1/2$, and, for the most part, turn out to be limited by
energy relaxation, i. e. $T_{2}\approx 2T_{1}$ (Fig.~\ref{Figure3}a).
Therefore, the noise that causes Ramsey fringes to decay is slow on the time
scale of order $T_{2}$ (about $10~\mathrm{\mu }$\textrm{s}) but fast on the
time scale of the typical Ramsey fringe acquisition time, of order one
minute, typical of $e$-jump rates seen with superconducting single electron
transistors and charge qubits \cite{Aumentado2004, Houck}.

\textit{Experimental evidence for the CQPS.} Analyzing our time-domain data,
we find that the measured flux dependence of the decoherence times $%
T_{2}^{\ast }(m_{\mathrm{ext}})$ of the $g\leftrightarrow e$ transition is
in excellent agreement with the expressions~(\ref{A-Clinewidth}) and (\ref%
{A-Clinewidth2}). Moreover, our data is inconsistent with conventional
decoherence mechanisms, which include low-frequency noise in the flux $m_{%
\mathrm{ext}}$, the Josephson energy $E_{J}$ (due to critical current noise
of the black-sheep junction), and in the inductive energy $E_{L}$ (due to
critical current noise in array junctions).

For the flux-noise mechanism, according to~Eq.~(\ref{FluxNoise}) and Fig.~%
\ref{Figure2}, we expect $T_{2}^{\ast }(m_{\mathrm{ext}})$ to be maximal at $%
|m_{\mathrm{ext}}|=1/2$ (flux sweet-spot). This is obviously inconsistent
with our data, which shows a clear minimum of $T_{2}^{\ast }(m_{\mathrm{ext}%
})$ at $|m_{\mathrm{ext}}|=1/2$, see Fig.~\ref{Figure3}a. Dephasing due to
fluctuations in the array inductance, given by Eq.~(\ref{ELnoise}), is also
ruled out because the corresponding $T_{2}^{\ast }(m_{\mathrm{ext}})$ would
also peak at $|m_{\mathrm{ext}}|=1/2$. (Qualitaively, this is because at $%
|m_{\mathrm{ext}}|=1/2$ the transition frequency $\nu _{eg}$ is given by $%
|E_{S}^{\Sigma }|/h$, which is a property of the black-sheep junction, and
is independent of the array inductance~\cite{ESnote}.) The numerically
evaluated sensitivity of the transition frequency $\nu _{eg}$ to $E_{L}$ is
presented in Fig.~\ref{FIG-Sensitivities}b. Theoretical analysis of the
sensitivity of $\nu _{eg}$ to the critical current noise in the black-sheep
junction (Eq.~(\ref{EJnoise}), Fig.~\ref{FIG-Sensitivities}a) shows that the
sensitivity must turn zero at some device-specific flux bias. Such
non-monotonic dependence is inconsistent with the observed monotonic $%
T_{2}^{\ast }(m_{\mathrm{ext}})$.

Our key result is that the highly unusual flux-dependence of the dephasing
time $T_{2}^{\ast }(m_{\mathrm{ext}})$ (Fig.~\ref{Figure3}a) is well
explained with the Aharonov-Casher effect of phase-slip interference. Upon
adding the measured small contribution $1/2T_{1}$ to the expression (\ref%
{CQPSdephasingTheory}) obtained from our theory of CQPS-induced dephasing,
see Eqs.~(\ref{MainHamiltonianPart1})-(\ref{A-Clinewidth2}) and Appendix~\ref%
{ACbroadening}, we find excellent agreement with the data (Fig.~\ref{Figure3}%
a). The only adjustable parameter we use is the theoretical value of $%
T_{2}^{\ast }$ at $|m_{\mathrm{ext}}|=1/2$. Furthermore, this adjustable
parameter agrees well with a WKB\ calculation of $E_{SA}\approx h\times 150$~%
\textrm{kHz} based on our estimates of the array junction parameters.

\textit{Control experiments.} To verify our interpretation of the dephasing
data in terms of the CQPS-induced broadening, we performed two main control
experiments.

In the first control experiment, we check if phase-slips interference can
explain the dephasing of some other fluxonium transition. For such
experiment it is necessary to select a transition other than the lowest $%
g\leftrightarrow e$, but with sufficiently long $T_{1}$, so that the
decoherence is dominated by dephasing and not by energy relaxation. A good
candidate for this control experiment turns out the transition to the second
excited state $g\leftrightarrow f$ in a narrow vicinity of $m_{\mathrm{ext}%
}=0$ (Fig.~\ref{Figure2}c). Interestingly, for this transition the flux bias 
$m_{\mathrm{ext}}=0$ turns out to be a sweet-spot for $T_{1}$, see Fig.~\ref%
{Figure3}c. The sharp increase of $T_{1}$ of the $g\leftrightarrow f$
transition at $m_{\mathrm{ext}}=0$ is well explained, see Eq.~(\ref%
{T1threelevel}), by exactly the same model of dissipation, $R_{0}(\omega
)=A/\omega $, used to explain energy relaxation of the $g\leftrightarrow e$
transition, with $A=190~M\Omega \times \mathrm{GHz}$. The origin of the peak
in $T_{1}$ at $m_{\mathrm{ext}}=0$ comes from the fact that the $f$ state
can decay either directly to $g$ or first to $e$ and then to $g$\textbf{; }at%
\textbf{\ }$m_{\mathrm{ext}}=0$, due to parity conservation, the direct $%
f\rightarrow g$ decay is forbidden~\cite{Manucharyan2009}. Thus, the
bottleneck at $m_{\mathrm{ext}}=0$ is a $280~\mathrm{MHz}$ $f\leftrightarrow
e$ transition resulting in the lifetimes of the $g\leftrightarrow f$
transition of up to $3~\mu s$.

The Ramsey fringe measurement on the $g\leftrightarrow f$ transition at $m_{%
\mathrm{ext}}=0$, yields $T_{2}^{\ast }\simeq 2.5$~$\mu s$, a significantly
smaller value than $2T_{1}\simeq 6~\mu s$. Echo measurement at $m_{\mathrm{%
ext}}=0$ yields a value of $T_{2}$ almost matching $2T_{1}$, see Fig.~\ref%
{Figure3}c. Thus, there is a noticeable amount of inhomogeneous broadening
of the $g\leftrightarrow f$ transition at $m_{\mathrm{ext}}=0$. We find that
this broadening is indeed precisely accounted for by expressions~(\ref%
{A-Clinewidth}), (\ref{A-Clinewidth2}), and (\ref{CQPSdephasingTheory}). In
other words, the ratio of $T_{2}^{\ast }=250$~ns of the $g\leftrightarrow e$
transition at $|m_{\mathrm{ext}}|=1/2$ and $T_{2}^{\ast }=2.5$~$\mu $s of
the $g\leftrightarrow f$ transition at $m_{\mathrm{ext}}=0$ coincides with
the theoretical prediction~$F_{gf}(0)/F_{ge}(1/2)$, where the overlap
integrals are computed without any adjustable parameters. The reduction of $%
T_{2}^{\ast }$ with the departure from the sweet-spot $m_{\mathrm{ext}}=0$
is also correctly reproduced without adjustable parameters~(Fig.~\ref%
{Figure3}c).

The second control experiment is even more telling, and is performed on
another fluxonium device, according to the following logic. Our theory of
dephasing by the interfereing quantum phase-slips predicts that increasing
the Josephson energy $E_{J_{j\neq 0}}$\ of the array junctions while leaving
their charging energy $E_{C_{j\neq 0}}$\ the same, would suppress
exponentially\cite{ESnote} the phase-slip matrix element $E_{SA}$. (Let us
remind that we assume all the junctions of the fluxonium array to be nearly
identical, i.e. $E_{J_{i}}\simeq E_{J_{j}}$\ and $E_{C_{i}}\simeq E_{J_{j}}$%
\ for any $i,~j>0$,\ so that $|E_{S_{i}}|\simeq |E_{S_{j}}|=E_{SA}$.) Thus,
we shall be able to practically switch off the decoherence of fluxoium
transitions due to CQPS in its array inductance, by a small adjustment of
the array junctions parameters.

\begin{figure}[tbp]
\centering
\resizebox{\linewidth}{!}{\includegraphics{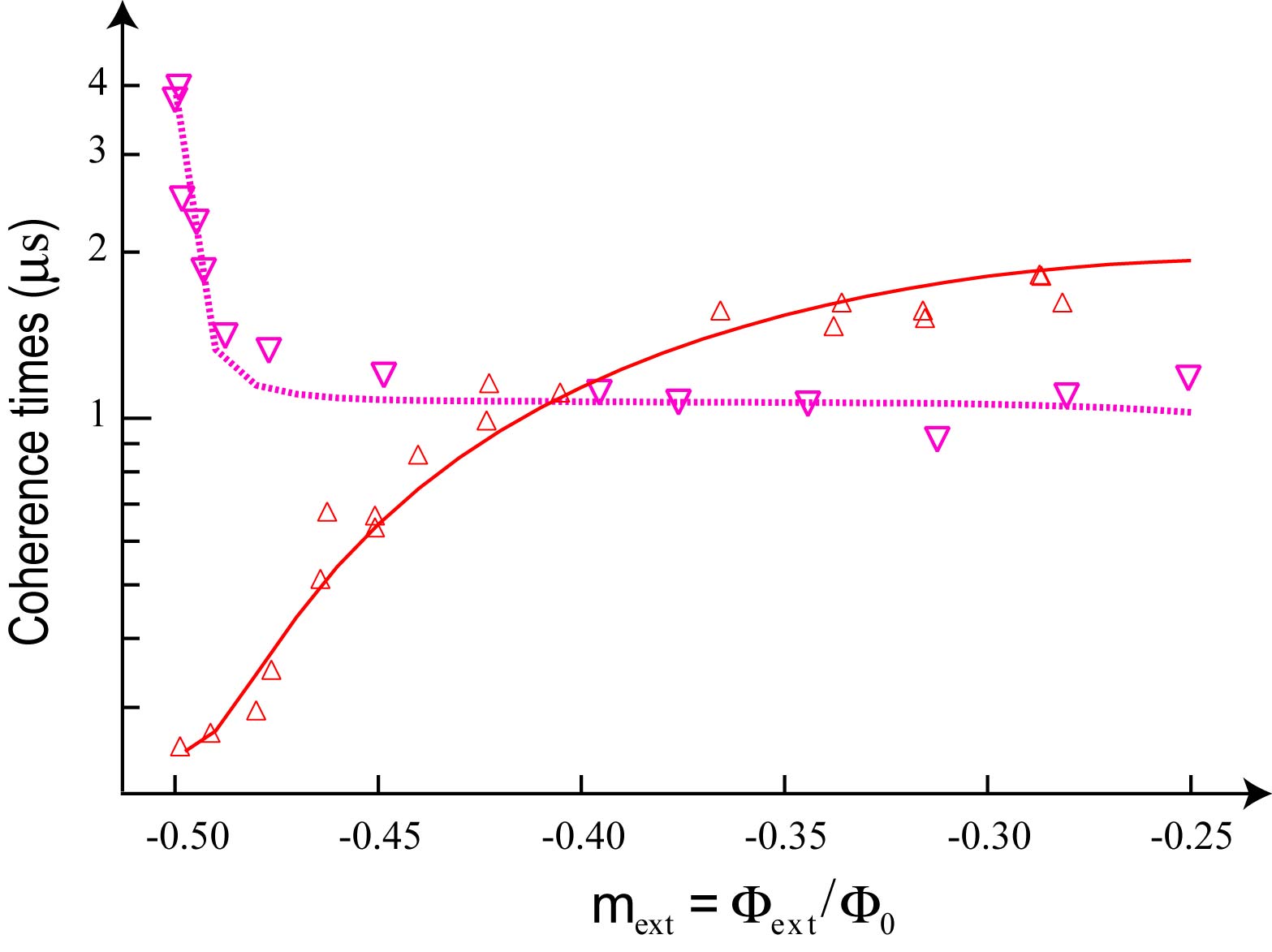}}
\caption{Measurement of the dephasing times $T_{2}^{\ast }$ of a new
fluxonium device with supressed CQPS rate (magenta inverted triangles) and
theoretical prediction (magenta dotted line). Data from Fig. 3a of the main
text (red triangles) on $T_{2}^{\ast }$ of the main device is shown for easy
comparison.For $|m_{\mathrm{ext}}|<0.25$, energy relaxation dominates the
decoherence.}
\label{Device2}
\end{figure}

Parameters of the fluxonium device for the second control experiment were
extracted from spectroscopy data (not shown) to be $E_{J_{j=0}}\equiv
E_{J}=12.0~\mathrm{GHz}$, $E_{C_{j=0}}\equiv E_{C}=2.46~\mathrm{GHz}$, $%
E_{L}=0.89~\mathrm{GHz}$, very similar to the parameters of the main
device~(see Table~\ref{Table1}). Since we kept the geometry of the array
junctions and their number identical to the previous device, we infer that
the charging energy of the array junctions $E_{C_{j\neq 0}}$\ did not change
significantly, while the Josephson energy $E_{J_{j\neq 0}}=N\times
E_{L}=43\times 0.89=38~\mathrm{GHz}$\ increased by a factor of $1.7$. For
such device parameters, the spectrum of this fluxonium is similar to that of
the main one. However, in sharp contrast to the main device, we find that $%
T_{2}^{\ast }(m_{\mathrm{ext}})$\ for the $g\leftrightarrow e$\ transition
is now sharply peaked at $|m_{\mathrm{ext}}|=1/2$\ and is nearly independent
on $m_{\mathrm{ext}}$\ away from this spot (Fig.~\ref{Device2}). The
reduction of $T_{2}^{\ast }$\ away from the sweet-spot follows the
prediction of the first-order flux-noise effect, Eq.~(\ref{FluxNoise}),
assuming the flux noise amplitude $\delta m_{\mathrm{ext}}\approx 10^{-5}$.
The dependence $T_{2}^{\ast }(m_{\mathrm{ext}})$\ away from the spot $|m_{%
\mathrm{ext}}|=1/2$\ is now clearly inconsistent with the CQPS. (They still
may be responsible for the dephasing at $|m_{\mathrm{ext}}|=1/2$.) The
measured value of $T_{2}^{\ast }=4~\mathrm{\mu s}$\ exactly at $|m_{\mathrm{%
ext}}|=1/2$\ is too small to be explained by the second order flux-noise,
but matches, within the factor of two, the prediction of Eq.~(\ref%
{A-Clinewidth}) for the CQPS-induced dephasing. Thus, a second fluxonium
device with nearly similar parameters, apart from an increase in $%
E_{J_{j\neq 0}}$\ by a factor of $1.7$\ resulted in a $16$-fold enhancement
of the coherence time at $|m_{\mathrm{ext}}|=1/2$, indicating the expected%
\cite{ESnote} exponential suppression of the phase-slip interference.

\section{Summary}

Coherent quantum phase-slips hybridize states of a superconductor with
different configurations of the order parameter. The corresponding
hybridization energy is exponentially small for \textquotedblleft
good\textquotedblright\ superconductors. In our experiment this energy, $%
|E_{S}|$, corresponds to only a few hundred $\mathrm{kHz}$, three orders of
magnitude lower than the $15~\mathrm{mK}$ sample temperature, and five
orders of magnitude lower than the main qubit energy scales, i. g. plasma
frequency and the inductive energy (both of order $10~\mathrm{GHz}$).
Detection of the tiny energy scale $|E_{S}|$ has been made possible by two
ingredients specific to our experiment. First, the Aharonov-Casher
modulation of CQPS amplitude broadens the qubit transitions. Second, the the
immunity of the fluxonium circuit to dissipation ($Q=\nu _{eg}T_{1}\gtrsim
10^{5}$) allows us the high-precision measurement of this broadening, thus
revealing CQPS. By replacing the junction array of the present experiment
with an amorphous superconducting nanowire, but keeping the black-sheep
junction, one may attempt seeing CQPS in nominally continuous wires with
poorly controlled CQPS\ amplitudes.

Our experiment also shows that the fluxonium artificial atom may find
applications in various quantum information processing schemes: it provides
a 3-level system displaying a combination of larger frequency range and
anharmonicity than most other qubits; it can operate away from flux sweet
spots without loosing too much coherence; its coupling to a microwave cavity
can be varied from weak to strong for exchange of quantum information
without the side effect of spontaneous emission. In the event that critical
current noise would end up dominating superconducting qubit coherence, one
may expect a $1/\sqrt{N}$ suppression of this effect using an $N$ junction
array. Finally, demonstrated coherence quality factor $Q$ exceeding $10^{5}$
in a circuit containing as many as $44$ junctions encourages the design of
large quantum Josephson networks, especially those offering topologically
protected ground states.

\acknowledgments

We thank M. Brink, C. Rigetti, D. Schuster, L. DiCarlo, J. Chow, L. Bishop,
H. Paik, I. Protopopov, L. Frunzio, R. Schoelkopf and S. Girvin for useful
discussions. This research was supported by the NSF under grants
DMR-0653377, DMR-1006060; the NSA through ARO Grant No. W911NF-09-1-0514,
IARPA under ARO contract No. W911NF-09-1-0369, DOE contract No.
DE-FG02-08ER46482, the Keck foundation, and Agence Nationale pour la
Recherche under grant ANR07-CEXC-003. M.H.D. acknowledges partial support
from College de France.

\appendix

\section{Experimental techniques}

\subsection{Sample description}

\label{APP-ChipDescription}

\begin{figure*}[tbp]
\resizebox{0.8\linewidth}{!}{\includegraphics{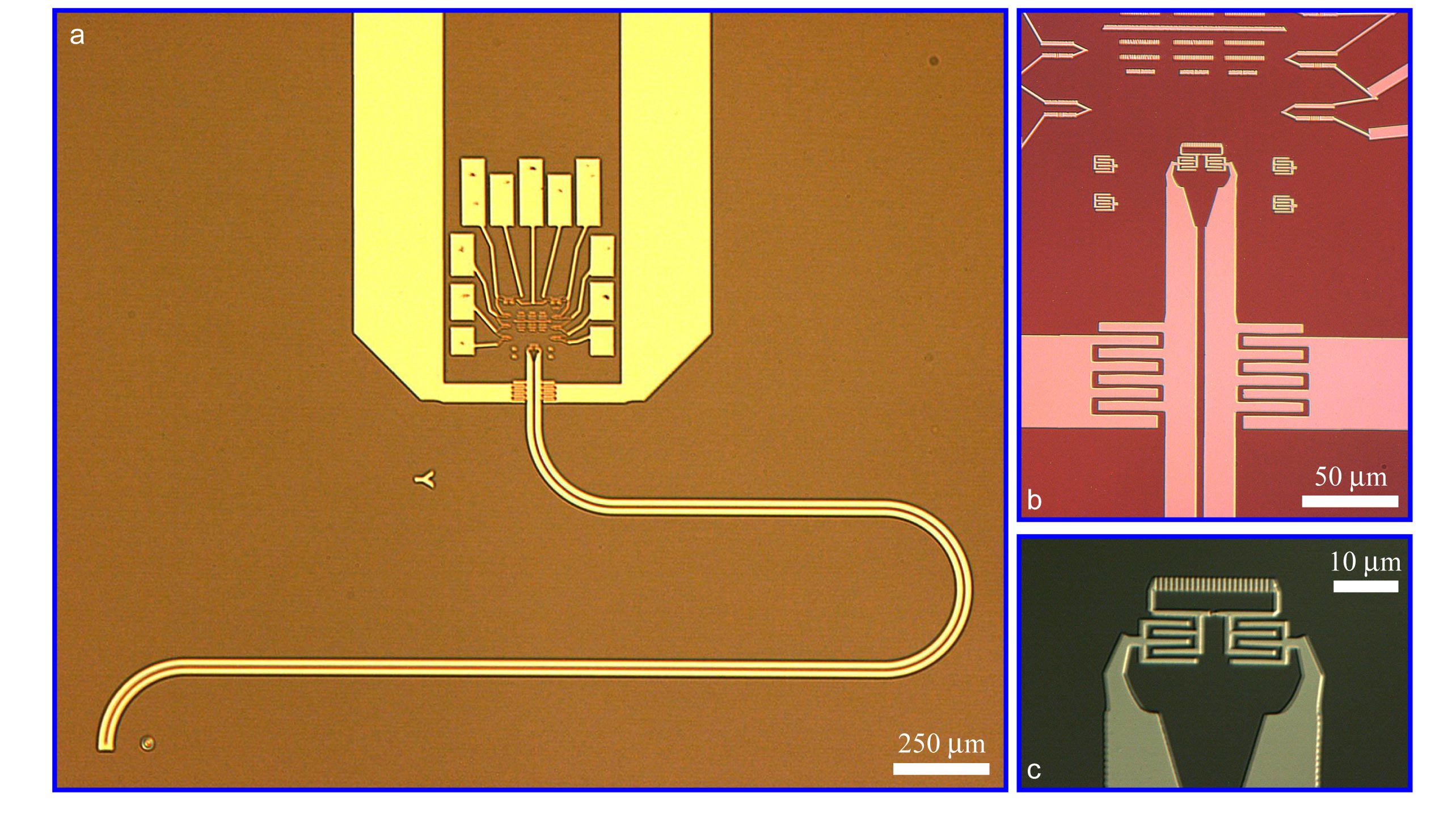}}
\caption{(a) Optical image of the sample (color filters applied for better
contrast), bright indicates Al, dark indicates Si substrate. Sample contains
the CPS resonator, the qubit, measurement leads and test structures. (b)
Zoom in on the voltage antinode region of the resonator. (c) Further zoom on
the fluxonium qubit loop.}
\label{FIG-DevicePhotograph}
\end{figure*}

The qubit and the readout circuits (Fig.~\ref{FIG-DevicePhotograph}a) are
fabricated on a Si chip using Al double-angle evaporation through a
suspended electron beam resist mask. The readout part consists of a
resonator, implemented with a $\lambda /4$ coplanar-strips (CPS)
transmission line. Resonator is coupled to $50~\Omega $ measurement leads
(microstrips) using a pair of interdigitated capacitors (Fig.~\ref%
{FIG-DevicePhotograph}b). The two leads of the \textquotedblleft
black-sheep\textquotedblright\ junction are connected to the resonator
strips with another pair of smaller interdigitated capacitors (Fig.~\ref%
{FIG-DevicePhotograph}c). A number of resistance and dose test structures
are placed outside the resonator.

The fabrication procedure is outlined in the previous work~\cite%
{Manucharyan2009}, with relevant device parameters collected into Table~(\ref%
{Table1}). We emphasize that the entire fluxonium device -- including the
Josephson array, the black-sheep\ junction, the resonator, and the test
structures -- is fabricated in a single step of e-beam lithography and
double-angle evaporation. Such simplification in the fabrication of a
superconducting qubit has been made possible because: i) the dimensions of
both the black-sheep\ junction and the array junctions are chosen to be
sufficiently close for patterning both types of junction in a single resist
mask ii) the strips of the CPS resonator are sufficiently narrow so that the
e-beam and lift off process used for the small junction fabrication could be
readily applied to the resonator; in addition, space remains to accommodate
test junctions and arrays fabricated simultaneously with the qubit.

\begin{table*}[h]
\begin{tabular}{|l||r|}
\hline
Readout resonator strips width (measured) & $15~\mathrm{\mu m}$ \\ \hline
Readout resonator strips separation (measured) & $4~\mathrm{\mu m}$ \\ \hline
Readout resonator wave impedance $Z_{\infty }$ (inferred) & $80~\Omega $ \\ 
\hline
Readout resonator resonance frequency $\nu _{0}$ (fit) & $8.175~\mathrm{GHz}$
\\ \hline
Readout resonator external quality factor $Q_{\mathrm{ext}}$ (measured) & $%
400$ \\ \hline
Readout resonator internal quality factor $Q_{\mathrm{int}}$ (measured) & $%
2000-5000$ \\ \hline
Black-sheep\ junction dimensions (nominal) & $0.35~\mathrm{\mu m\times }0.2~%
\mathrm{\mu m}$ \\ \hline
Array junction dimensions (nominal) & $2~\mathrm{\mu m\times }0.2~\mathrm{%
\mu m}$ \\ \hline
Array inductive energy $E_{L}=(\hbar /2e)^{2}/L$ (fit) & $0.525~\mathrm{GHz}$
\\ \hline
Array inductance $L$ (inferred) & $300~\mathrm{nH}$ \\ \hline
Black-sheep junction Josephson energy $E_{J_{j=0}}\equiv E_{J}$ (fit) & $8.9~%
\mathrm{GHz}$ \\ \hline
Black-sheep junction Coulomb energy $E_{C_{j=0}}\equiv E_{C}=e^{2}/2C_{J}$
(fit) & $2.5~\mathrm{GHz}$ \\ \hline
Number of array junctions $N$ (nominal) & $43$ \\ \hline
Array junction Josephson energy $E_{j\neq 0}=NE_{L}$ (inferred) & $22.5~%
\mathrm{GHz}$ \\ \hline
Array junction Coulomb energy $E_{C_{j\neq 0}}$ (inferred) & $0.85-1~\mathrm{%
GHz}$ \\ \hline
Qubit-cavity coupling constant $g$ (fit) & $181~\mathrm{MHz}$ \\ \hline
Qubit-cavity coupling capacitance $C_{c}$ (inferred from $g$ and $Z_{\infty
} $) & $0.8~\mathrm{fF}$ \\ \hline
Black-sheep\ junction phase-slip energy $|E_{SB}|$ (measured, inferred) & $%
369~\mathrm{MHz}$ \\ \hline
Array junction phase-slip energy $E_{SA}$ (inferred from $L$, $E_{J_{j\neq
0}}$, $E_{C_{j\neq 0}}$) & $50-250~\mathrm{kHz}$ \\ \hline
Array junction phase-slip energy $E_{SA}$ (inferred from $T_{2}^{\ast }(m_{%
\mathrm{ext}}=1/2)$) & $130~\mathrm{kHz}$ \\ \hline
\end{tabular}%
\label{Table1}
\caption{Device parameters}
\end{table*}

\subsection{Microwave reflectometry readout}

\label{APP-Reflectometry}

\begin{figure*}[tbp]
\resizebox{0.8\linewidth}{!}{\includegraphics{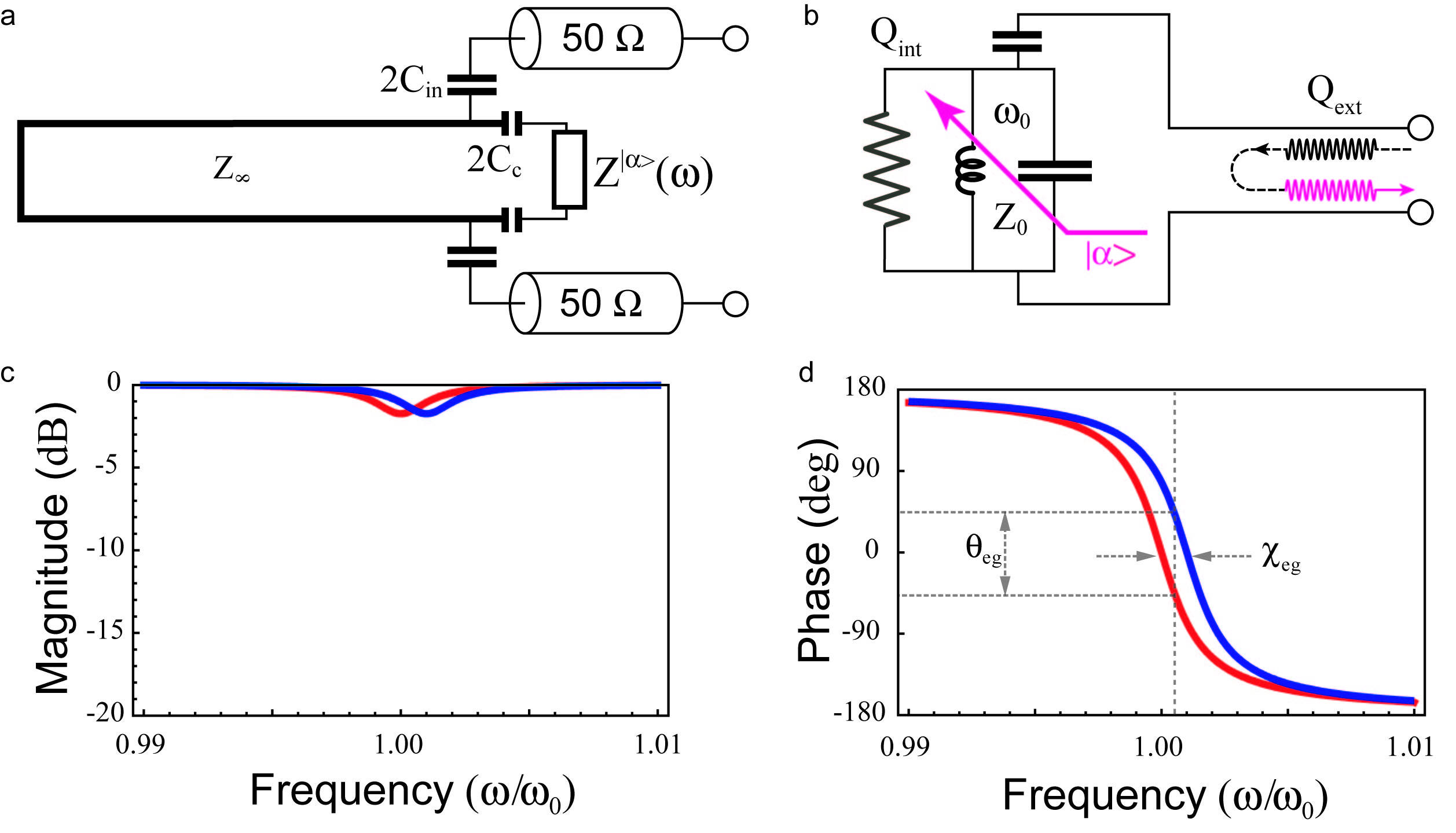}}
\caption{(a) Electrical model of a qubit interacting with the CPS
transmission line resonator. Thick solid line represents the distributed
transmission line. (b) Simplified circuit model of the dispersive effect of
qubit on resonator. (c) Modulus of the reflection amplitude in model of
panel (b) vs. signal frequency for $Q_{\mathrm{int}}=10Q_{\mathrm{ext}}$.
Blue and red traces correspond to qubit states $g$ and $e$ respectively. (d)
Phase of the reflection amplitude, same conditions as in panel (c).}
\label{cQEDschematic}
\end{figure*}

We use the lowest differential mode of our resonator, which corresponds to
the frequency $\omega _{0}$ at which the physical length of the CPS
transmission line matches a quarter of the wavelength. The next order
resonance lies at $3\omega _{0}$. The qubit can be viewed as a high
impedance termination $Z^{\left\vert \alpha \right\rangle }(\omega )$ placed
at the open end of the transmission line, and depends both on the frequency $%
\omega $ and the qubit state $\alpha $ (Fig.~\ref{cQEDschematic}a). It
provides a small contribution $\chi _{\alpha }$ - termed the dispersive
shift - to the resonance frequency $\omega _{0}$, when the qubit is in state 
$\left\vert \alpha \right\rangle $. The shift in the resonance frequency is
detected by monitoring the complex-valued reflection amplitude $\Gamma $ for
the scattering of a microwave signal off the resonator. Approximating the
single-mode resonance with an effective $LC$-oscillator (Fig.~\ref%
{cQEDschematic}b), the reflection amplitude is given by 
\begin{equation}
\Gamma (\omega ,\omega _{0})=\frac{2i(\frac{\omega -\omega _{0}}{\omega _{0}}%
)-Q_{\mathrm{ext}}^{-1}+Q_{\mathrm{int}}^{-1}}{2i(\frac{\omega -\omega _{0}}{%
\omega _{0}})+Q_{\mathrm{ext}}^{-1}+Q_{\mathrm{int}}^{-1}}~,
\label{ReflectionAmplitude}
\end{equation}%
where $Q_{\mathrm{ext}}$ is the quality factor due to the energy loss in the
matched $50~\Omega $ measurement leads while $Q_{\mathrm{int}}$ is the
quality factor due to the energy loss inside the resonator (electrically
represented by a resistor shunting the $LC$-circuit). In our CPS resonator $%
Q_{\mathrm{ext}}\simeq 400\ll Q_{\mathrm{int}}\approx 4000$ making $|\Gamma
(\omega ,\omega _{0})|$ to be very close to unity (Fig.~\ref{cQEDschematic}%
c). The phase $\theta =\mathrm{Arg}(\Gamma )$ of the reflected signal is a
rapid function of frequency: $\theta =2\arctan 2Q_{\mathrm{ext}}\frac{\omega
-\omega _{0}}{\omega _{0}}$ (Fig.~\ref{cQEDschematic}d). Finally, the
difference in phase of the reflection coefficient $\theta _{eg}=\theta
_{e}-\theta _{g}$ between the qubit excited state $e$ and the ground state $%
g $ is given by

\begin{equation}
\theta _{eg}=2\arctan \left( 2Q_{\mathrm{ext}}\frac{\chi _{g}}{\nu _{0}}%
\right) -2\arctan \left( 2Q_{\mathrm{ext}}\frac{\chi _{e}}{\nu _{0}}\right)
~,  \label{DispersivePhaseShift}
\end{equation}%
where $\nu _{0}=\omega _{0}/2\pi $ is the resonator frequency in $\mathrm{Hz}
$. The quantity $\theta _{eg}$ is plotted on the Y-axes of Fig.~\ref{Figure2}%
a-left, and plotted as the color scale in Fig.~\ref{Figure2}a-right and in
Fig.~\ref{Figure2}c.

Note that since $\theta _{eg}$ saturates quickly once $2Q_{\mathrm{ext}}%
\frac{\chi _{e}-\chi g}{\omega _{0}}>1$, increasing $Q_{\mathrm{ext}}$ above
the value $\omega _{0}/|\chi _{e}-\chi _{g}|$ will not improve the
sensitivity. In our experiment $|\chi _{e,f}-\chi _{g}|$ lie between $1~%
\mathrm{MHz}$ and $10~\mathrm{MHz}$, while $\nu _{0}=8.175~\mathrm{GHz}$, so 
$Q_{\mathrm{ext}}=400$ is a convenient choice with a measurement bandwidth
of $20~\mathrm{MHz}$. We discuss the origin of the dispersive shifts as well
as the effect of finite $Q_{\mathrm{ext}}$ on the qubit lifetimes (Purcell
effect) in the text below, see Eqs.~(\ref{Reff}), (\ref{T1Theory}), and (\ref%
{T1Purcell}).

\subsection{Measurement schematic}

\label{APP-MeasurementSchematic}

\begin{figure*}[tbp]
\centering\resizebox{0.8\linewidth}{!}{\includegraphics{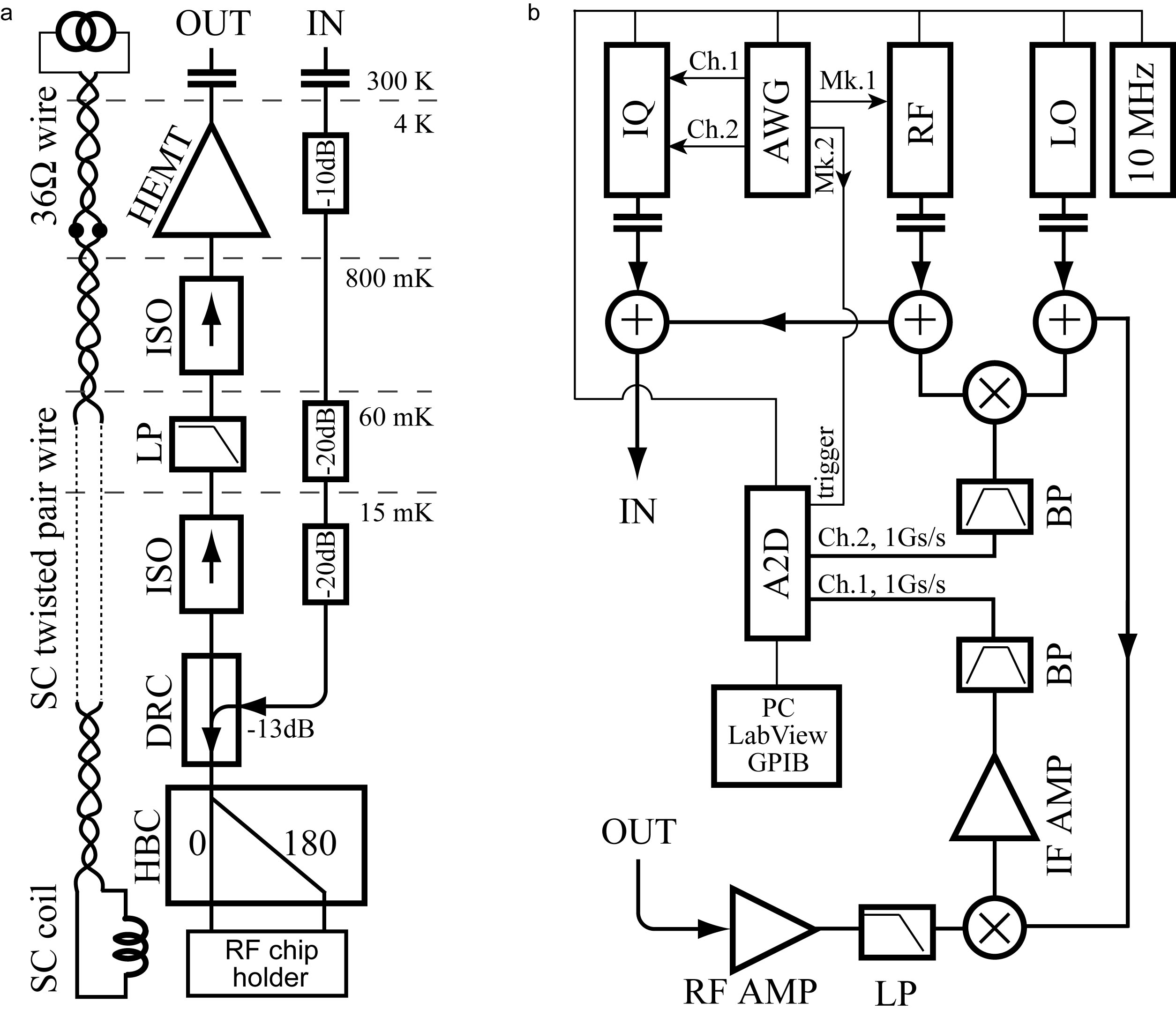}}
\caption{(a) Dilution refridgerator part of the microwave reflectometry
setup. HEMT - high electron mobility cryogenic amplifier, ISO - cryogenic
broadband isolator, LP - low pass filter, DRC - directional coupler, HBC -
180 degree hybrid coupler, see text for more details (b) Room-temperature
signal generation and demodulation setup. LO - continuous microwave source,
RF - pulsed (Mk.1 controlled) wave source, IQ is a vector signal generator
for driving the qubit transitions, AWG - arbitrary waveform generator, A2D-
two-channel fast digitizer, \textquotedblleft $+$\textquotedblright\ -
matched signal combiner/splitter, \textquotedblleft $\times $%
\textquotedblright\ - microwave mixer, BP- and pass filter,
\textquotedblleft $10~\mathrm{MHz}$\textquotedblright\ - Rb reference, RF
AMP - a room temperature microwave amplifier.}
\label{FIG-MeasurementSetup}
\end{figure*}

\begin{figure*}[tbp]
\resizebox{0.8\linewidth}{!}{\includegraphics{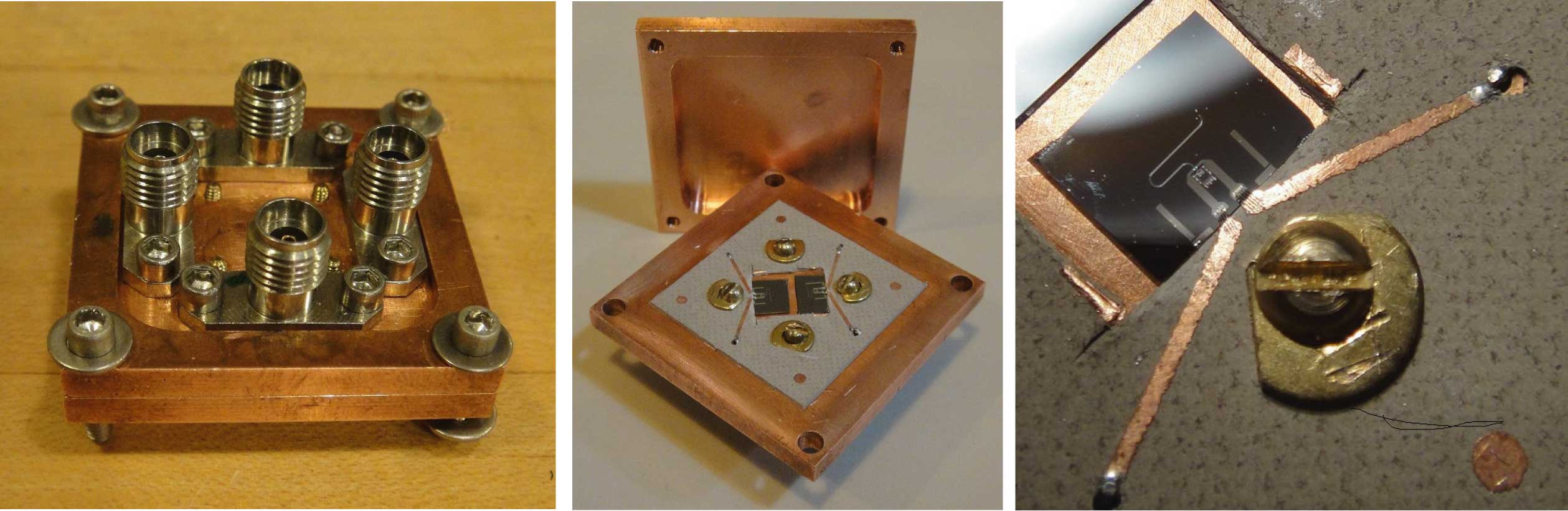}}
\caption{Photographs of the chip holder carrying the actual fluxonium
device. Anritsu K connector guides microwaves from coaxial cables to Cu
microstrip lines ($300~\mathrm{\protect\mu m}$ wide) on the printed circuit
board. The center conductor of the K connector is soldered to the begining
of the microstrip. Microstrip lines continue to the chip by means of 5-6
short wirebonds.}
\label{FIG-SampleHolder}
\end{figure*}

\textit{Low-temperature setup }(Fig.~\ref{FIG-MeasurementSetup}a)\textit{.}
The incoming signals line is attenuated with cryogenic high-frequency
resistive film attenuators (XMA), with total attenuation exceeding $50~%
\mathrm{dB}$. The readout line is shielded by the two $4-12~\mathrm{GHz}$
isolators (Pamtech) and a low-pass filter (K\&L) with rejection band $10-40~%
\mathrm{GHz}$. Outgoing signals are amplified using a $5~\mathrm{K}$ noise
temperature cryogenic HEMT amplifier (Caltech). Incoming and outgoing waves
are separated from each other using a directional coupler (Krytar).
Differential excitation of the resonator is implemented using a $180$ degree
hybrid coupler (Krytar). The chip rests at the bottom of the fully enclosing
copper sample holder, the microwaves are guided to it by means of two
printed circuit board microstrips wirebonded to their on-chip continuation,
and perpendicular coaxial-to-microstrip transition implemented using Anritsu
K connectors (Fig.~\ref{FIG-SampleHolder}). The transition provides less
than $15~\mathrm{dB}$ return loss up to $20~\mathrm{GHz}$. Flux bias is
provided by driving a hand-made superconducting coil (glued to the sample
holder) with DC currents of order $1-10~\mathrm{mA}$.

\textit{Room-temperature setup }(Fig.~\ref{FIG-MeasurementSetup}b)\textit{. }%
The readout signal is provided with Agilent E8257D generator (RF), the qubit
pulses are generated using Agilent E8267D vector signal generator (IQ)
combined with Tektronix 520 arbitrary waveform generator (AWG). Both readout
and qubit signals are combined at room temperature and sent into the IN line
of the refrigerator. The reflected $\sim 8~\mathrm{GHz~}$readout signal from
the refrigerator OUT line is amplified at room temperature with a Miteq
amplifier ($1-12~\mathrm{GHz}$, $30~\mathrm{dB}$ gain), mixed down with a
local oscillator signal (LO), provided by HP 8672A, to a $50~\mathrm{MHz}$
IF signal, then filtered and amplified with the IF amplifier (SRS SR445A),
and finally digitized using one channel of the $1~\mathrm{GS}/\mathrm{s}$
Agilent Acqiris digitizer. A reference IF signal is created by mixing a copy
of RF and LO and digitized using the second channel. A software procedure
then subtracts the phases of the two IF signals, resulting in a good
long-term stability of the phase measurement. The short-term stability is
implemented by phase locking every instrument to a Rb $10~\mathrm{MHz}$
reference (SRS FS725). The marker signals of the AWG are used as triggers to
other instruments. Typical time to acquire Ramsey fringes of $5~\mathrm{\mu s%
}$ long (Fig.~\ref{Figure3}b) ranges from $10$ seconds to $1$ minute,
without noticeable change in the fringe decay time. The magnetic coil is
biased with Yokogawa 7751 voltage source in series with a $1:10$ voltage
divider and a $1~\mathrm{k\Omega }$ resistor at room temperature.

\begin{figure*}[tbp]
\resizebox{0.8\linewidth}{!}{\includegraphics{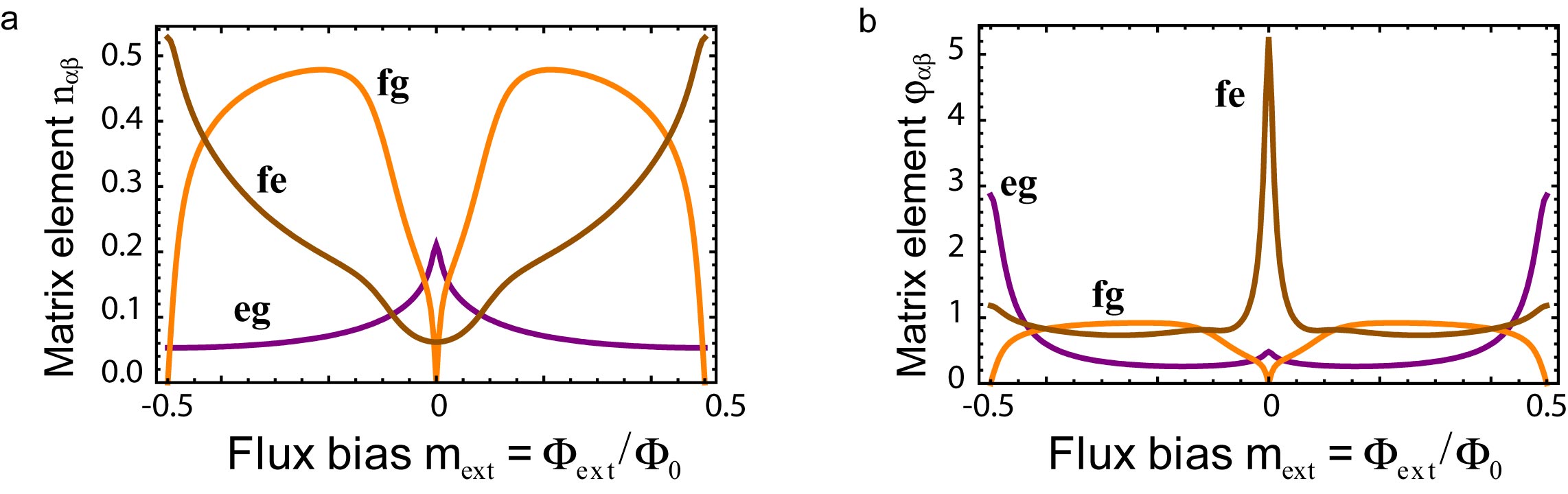}}
\caption{{}(a) Matrix elements of charge operator $N$ (in units of Cooper
pair charge $2e$) for transitions $g\leftrightarrow e$, $g\leftrightarrow f$%
, and $e\leftrightarrow f$. (b) Matrix elements of phase operator (in
radians) for same transitions.}
\label{FIG-MatrixElements}
\end{figure*}

\subsection{Dispersive shifts of a resonator by fluxonium artificial atom}

\label{APP-DispersiveShifts}

Here we evaluate $\chi _{\alpha }$ used in Eq.~(\ref{DispersivePhaseShift})
for the reflectometry signal (Fig.~\ref{cQEDschematic}d). The interaction of
the qubit with the resonator is given by

\begin{equation}
\hat{V}=hg(\hat{a}+\hat{a}^{\dag })\hat{n}\,,
\end{equation}%
where $\hat{n}$ is the charge on black-sheep\ junction capacitor in units of 
$2e$, and $\hat{a}$ is the annihilation operator for the microwave photons
in the equivalent $LC$-oscillator (Fig.~\ref{cQEDschematic}a-b). The
coupling constant $g$, given by 
\begin{equation}
g=\frac{C_{c}}{C_{J}+C_{c}}\sqrt{\frac{1}{2}Z_{0}/R_{Q}}\nu _{0}~,
\end{equation}%
is expressed via the characteristic impedance of the oscillator $%
Z_{0}=4Z_{\infty }/\pi $ ($Z_{\infty }=70-80~\Omega $ is the wave impedance
of the CPS transmission line), and the superconducting impedance quantum $%
R_{Q}=\hbar /(2e)^{2}\simeq 1~\mathrm{k\Omega }$. Treating $\hat{V}$ as a
perturbation to second order in $g$ yields the following expression for the
dispersive shift $\chi _{a}$ of the resonator frequency for the qubit in
state $\left\vert \alpha \right\rangle $: 
\begin{equation}
\chi _{\alpha }=2g^{2}\sum_{\beta \neq \alpha }\frac{|n_{\alpha \beta
}|^{2}\nu _{\alpha \beta }}{\nu _{\alpha \beta }^{2}-\nu _{0}^{2}}\,,
\label{DispersiveShift}
\end{equation}%
where $n_{\alpha \beta }$ are the matrix element of the charge operator $%
\hat{n}$ connecting states $\alpha $ and $\beta $ and $\nu _{\alpha \beta }$
is the qubit transition frequency (in \textrm{Hz}) between these states. In
formula~\ref{DispersiveShift}, $\nu _{\alpha \beta }$ is taken negative if
the state $\alpha $ is higher in energy than the state $\beta $, and
positive otherwise; in the rest of the text, for simplicity, we treat $\nu
_{\alpha \beta }$ as a positive number. Perturbation theory breaks down
whenever the denominator goes to zero, a situation which corresponds to a
resonance between various qubit transitions and the resonator frequency. The
inset of Fig.~\ref{Figure2}c shows an uncommon instance of such vacuum Rabi
resonance with the qubit transition $e\leftrightarrow f$. A conventional
vacuum Rabi resonance, involving the lowest $g\leftrightarrow e$ qubit
transition, takes place at $|m_{\mathrm{ext}}|\simeq 0.05$ and $\nu
_{eg}\simeq 8.2~\mathrm{GHz}$, it is shown as an anticrossing of the theory
lines (actual data available elsewhere~\cite{Manucharyan2009}).

Interestingly, for the case of the present fluxonium artificial atom, the
dominant contribution to the dispersive shift of the lowest transition ($%
g\leftrightarrow e$) comes from the transitions $g\leftrightarrow f$ and $%
e\leftrightarrow f$ and not from the $g\leftrightarrow e$ transition itself.
This happens because the transitions to the $f$ state involve large charge
motion and the $\nu _{ef}$ frequency remains in a window $1-2~\mathrm{GHz}$
away from the cavity frequency $\nu _{0}$ for all flux values. This behavior
is due to a large participation of the black-sheep\ plasma mode in the $f$
state for the present device parameters. By contrast, the $g\leftrightarrow
e $ transition involves little charge motion (Fig.~\ref{FIG-MatrixElements})
because it connects states with different phase-slips number\textbf{\ }$m$,
and, in addition, detunes quickly from the cavity with the external flux.
The large splitting between the transition $e\leftrightarrow f$ and the
resonator shown in the inset of Fig.~\ref{Figure2}c of the main text
illustrates this point.

\section{Decoherence of fluxonium}

\subsection{Aharonov-Casher Line Broadening}

\label{ACbroadening} 
\begin{figure*}[tbp]
\resizebox{0.8\linewidth}{!}{\includegraphics{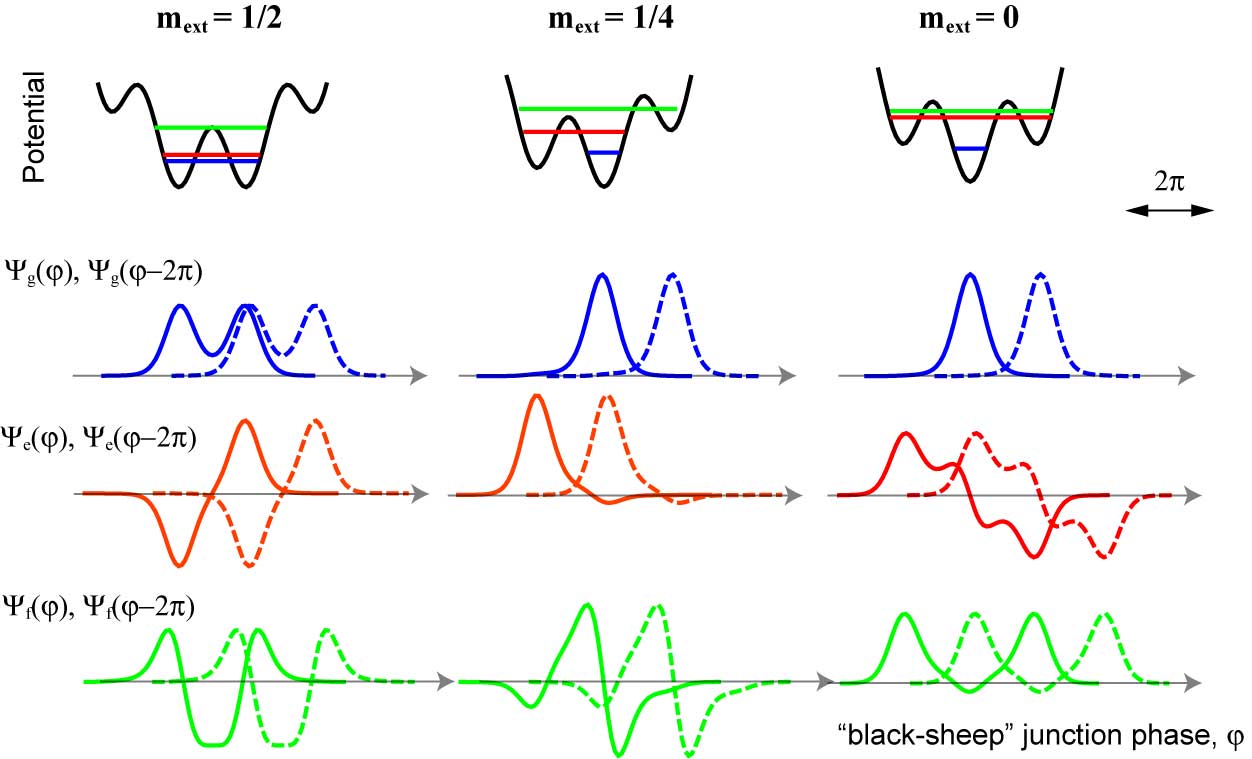}}
\caption{Top row: Effective potential for the \textquotedblleft
black-sheep\textquotedblright\ junction phase and a sketch of the three
lowest levels for three values of $m_{\mathrm{ext}}$. Bottom rows: Overlap
between $2\protect\pi $-shifted fluxonium wavefunctions in states $g$, $e$,
and $f$, for the three values of $m_{\mathrm{ext}}$. Note that for states $g$
and $e$ the overlap is clearly maximal when $|m_{\mathrm{ext}}|=1/2$.}
\label{FIG-WafeFunctionOverlap}
\end{figure*}

\begin{figure}[tbp]
\resizebox{0.8\linewidth}{!}{\includegraphics{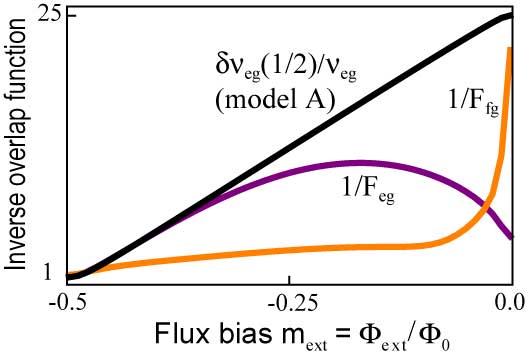}}
\caption{Flux-dependent part of the inverse linewidth due to coherent
quantum phase-slips and the Aharonov-Casher effect in the array. Black line
corresponds to the model A prediction for the $g\leftrightarrow e$
transition, purple and yellow lines present results of model B for the $%
g\leftrightarrow e$ and $g\leftrightarrow f$ transitions, respectively. Note
that close to $|m_{\mathrm{ext}}|=1/2$ the two models provide the same
result for the lowest transition.}
\label{FIG-ACdephasingTheory}
\end{figure}

Here we derive and discuss Eq.~(\ref{A-Clinewidth}). The Hamiltonian of
model B, defined by Eqs.~(\ref{MainHamiltonianPart1}) and (\ref%
{MainHamiltonianPart2}) is invariant under the transformation $(\varphi ,~%
\tilde{m})\rightarrow (\varphi -2\pi ,~\tilde{m}+1)$. This symmetry
represents the fact that by looking only at the initial and final states of
the junction loop we cannot distinguish which part of the loop (black-sheep
junction or the array) actually underwent a phase-slip. This point can also
be illustrated by representing graphically the phase distribution of array
islands before and after a phase-slip by $2\pi $(Fig.~\ref{Figure1}e).%
\textbf{\ }The unperturbed eigenstates $\left\vert \alpha \right\rangle
^{(0)}$ of the Hamiltonian (\ref{MainHamiltonianPart1}) then take the
following form: 
\begin{equation}
\left\vert \alpha \right\rangle ^{(0)}=\lim_{M\rightarrow \infty }\frac{1}{%
\sqrt{2M+1}}\sum_{\tilde{m}=-M}^{M}\Psi _{a}(\varphi -2\pi \tilde{m}%
)\left\vert \tilde{m}\right\rangle ~.
\end{equation}%
Here $\Psi _{\alpha }(\varphi )$ is the wavefunction of the $\alpha $-th
(non-degenerate) eigenstate of the fluxonium Hamiltonian Eq.~(\ref%
{MainHamiltonianPart2}), the states $\left\vert \tilde{m}\right\rangle $ are
the eigenstates of the integer $\tilde{m}$ operator, and the normalization
is chosen to satisfy $\left\langle \alpha |\beta \right\rangle ^{(0)}=\delta
_{\alpha \beta }$. Now, treating the quantum phase-slips perturbation
(second term of Eq.~(\ref{MainHamiltonianPart1})) to the first order in $%
|E_{S}|$, we find that the correction to the qubit transition frequency $\nu
_{\alpha \beta }^{\{1\}}$ between the states $\alpha $ and $\beta $:

\begin{multline*}
\nu _{\alpha \beta }^{\{1\}}=\frac{1}{2h}\lim_{M^{\prime },M^{\prime \prime
}\rightarrow \infty }\frac{1}{\sqrt{2M^{\prime }+1}\sqrt{2M^{\prime \prime
}+1}} \\
\times \sum\limits_{\tilde{m}=-\infty }^{+\infty }\sum\limits_{\tilde{m}%
^{\prime }=-M^{\prime }}^{M^{\prime }}\sum\limits_{\tilde{m}^{\prime \prime
}=-M^{\prime \prime }}^{M^{\prime \prime }}\int\limits_{-\infty }^{\infty }%
\mathrm{d}\varphi \\
\times (\Psi _{\alpha }(\varphi -2\pi \tilde{m}^{\prime })\Psi _{\alpha
}(\varphi -2\pi \tilde{m}^{\prime \prime }) \\
-\Psi _{\beta }(\varphi -2\pi \tilde{m}^{\prime })\Psi _{\beta }(\varphi
-2\pi \tilde{m}^{\prime \prime })) \\
\times \left( E_{S}\left\langle \tilde{m}^{\prime }|\tilde{m}\right\rangle
\left\langle \tilde{m}+1|\tilde{m}^{\prime \prime }\right\rangle
+E_{S}^{\ast }\left\langle \tilde{m}^{\prime }|\tilde{m}+1\right\rangle
\left\langle \tilde{m}|\tilde{m}^{\prime \prime }\right\rangle \right) ~.
\end{multline*}%
Because states with $\tilde{m}\neq \tilde{m}^{\prime }$ are orthogonal, the
sum reduces to a compact expression

\begin{equation}
\nu _{\alpha \beta }^{\{1\}}=\frac{\mathrm{Re}[E_{S}]}{h}F_{\alpha \beta
}(m_{\mathrm{ext}})~,  \label{Shift}
\end{equation}%
where the overlap function $F_{\alpha \beta }(m_{\mathrm{ext}})$ is defined
by Eq.~(\ref{A-Clinewidth2}). The flux dependence of $\nu _{\alpha \beta
}^{\{1\}}$ comes from the flux-dependence of the qubit state wavefunction,
several examples are illustrated in Fig.~\ref{FIG-WafeFunctionOverlap} and
in Fig.~\ref{FIG-ACdephasingTheory}.

In order to convert the shift $\nu _{\alpha \beta }^{\{1\}}$, Eq.~(\ref%
{Shift}), into the linewidth $\delta \nu _{\alpha \beta }$, Eq.~(\ref%
{A-Clinewidth}), let us recall that 
\begin{equation}
E_{S}=\sum\limits_{j=1}^{N}E_{S_{j}}\exp \left( i2\pi Q_{j}/2e\right) ~,
\end{equation}%
$\bigskip $with $Q_{j}$ being random variables with a spread of values
comparable to $e$, and the sum running over all array junctions. According
to the central-limit theorem, in the limit of large $N$, the quantity 
\textrm{Re}$[E_{S}]$ obeys to the gaussian distribution with zero mean{}and
standard deviation $\sigma =\sqrt{\overline{(\mathrm{Re}[E_{S}])^{2}}}$,

\begin{equation}
P\{0<\mathrm{Re}[E_{S}]<x\}=\frac{1}{\sqrt{2\pi }\sigma }\exp
(-x^{2}/2\sigma ^{2})~.  \label{Rayleigh}
\end{equation}%
Assuming the array junctions to be approximately identical, $E_{S_{j}}\simeq
E_{SA}$, we get $\overline{(\mathrm{Re}[E_{S}])^{2}}=\frac{1}{2}%
E_{SA}^{2}\times N$, and then readily compute the linewidth $\delta \nu
_{\alpha \beta }$ (defined as $\sqrt{\overline{\left( \nu _{\alpha \beta
}^{\{1\}}\right) ^{2}}}$) due to inhomogeneous broadening to be given by
Eq.~(\ref{A-Clinewidth}). If charges $Q_{j}$ vary slowly compared to the
duration of a single Ramsey fringe experiment (of order $10~\mathrm{\mu s}$%
), the decaying envelope of the of Ramsey fringes is given by the absolute
value of the characteristic function of the distribution (\ref{Rayleigh}).
We therefore find that Ramsey fringe envelope is given by a gaussian $\exp
\left( -\left( t/T_{\phi }^{\mathrm{CQPS}}\right) ^{2}\right) $, with the
flux-dependent dephasing time $T_{\phi }^{\mathrm{CQPS}}$ of the $\alpha
\rightarrow \beta $ transition due to the coherent quantum phase-slips given
by

\begin{equation}
1/T_{\phi }^{\mathrm{CQPS}}(m_{\mathrm{ext}})=\sqrt{2}\pi \delta \nu
_{\alpha \beta }(m_{\mathrm{ext}})~.  \label{CQPSdephasingTheory}
\end{equation}%
We used the expression (\ref{CQPSdephasingTheory}) to produce theory plots
in Fig.~\ref{Figure3}.

\subsection{Energy relaxation of fluxonium transitions ($T_{1}$ processes)}

\begin{figure}[tbp]
\centering
\resizebox{0.8\linewidth}{!}{\includegraphics{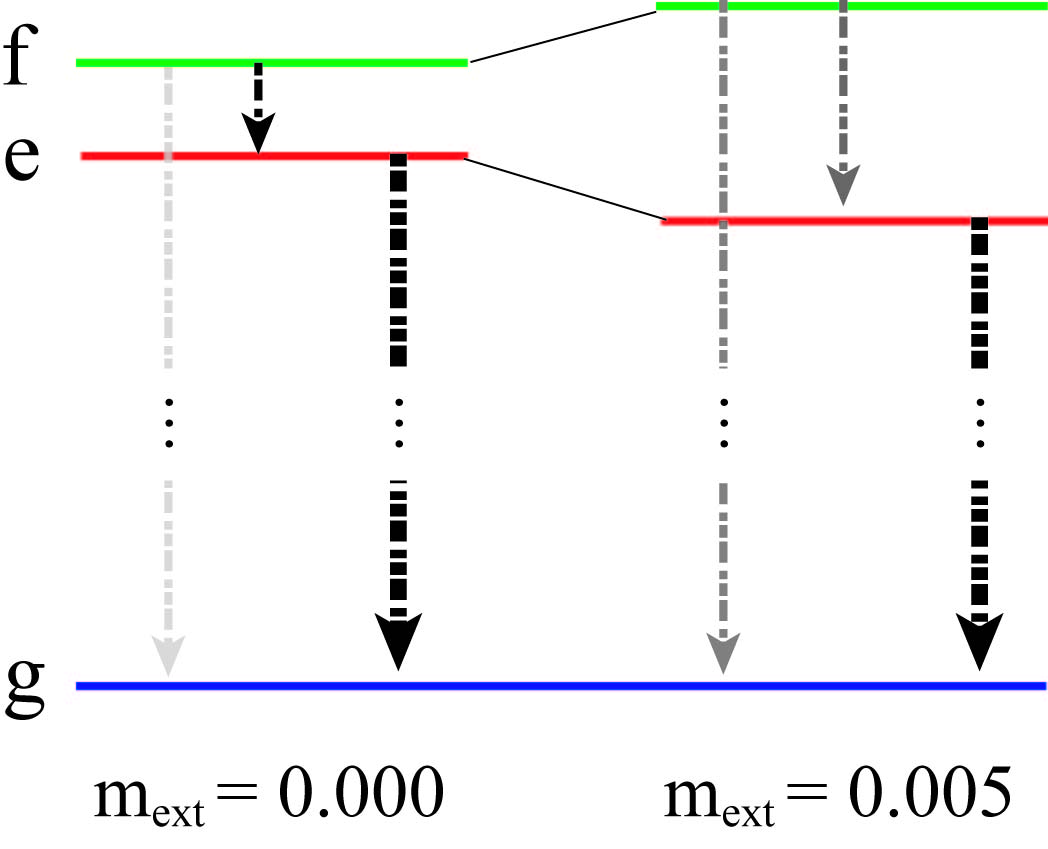}}
\caption{Sketch of the lowest three energy levels of fluxonium qubit and
structure of the corresponding relaxation rates in close proximity to $m_{%
\mathrm{ext}}=0$. Vanishing of the phase matrix element $\protect\varphi %
_{fg}$ at $m_{\mathrm{ext}}=0$ combined with the particular frequency
dependence of $R_{\mathrm{intrinsic}}(\protect\omega )$ explains the
sweet-spot in $T_{1}^{fg}$ at zero flux bias.}
\label{FIG-ThreeLevelRelaxation}
\end{figure}

\begin{figure*}[tbp]
\resizebox{0.8\linewidth}{!}{\includegraphics{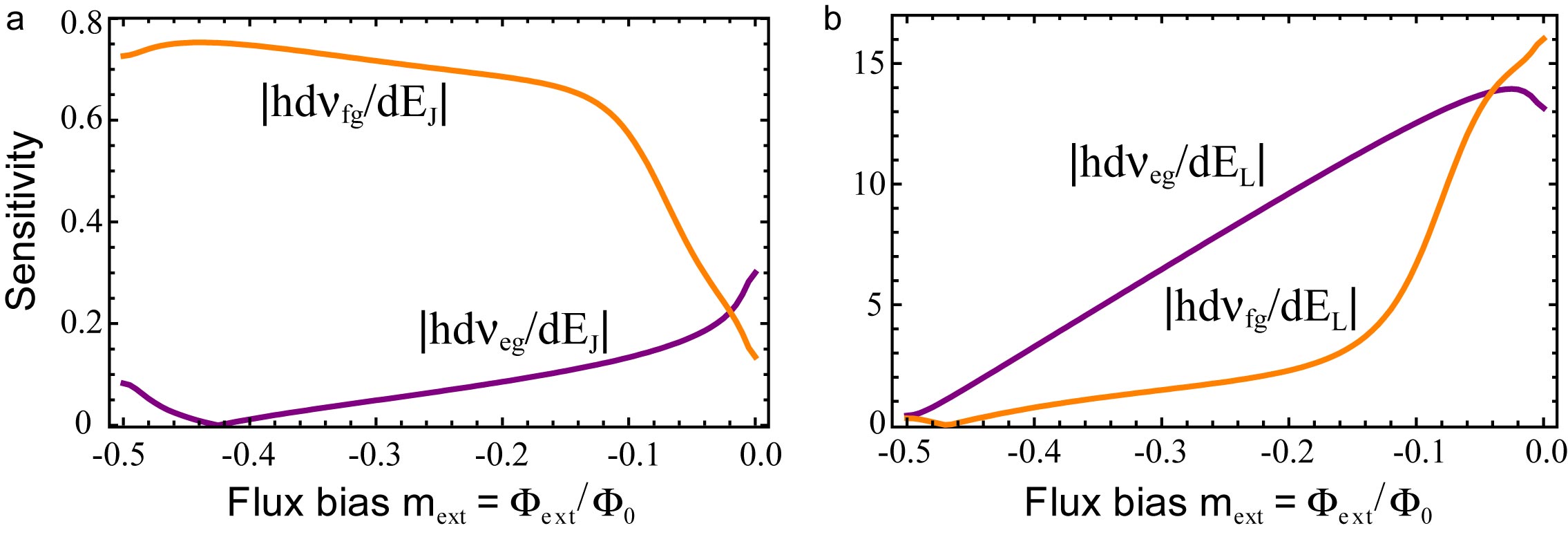}}
\caption{Sensitivity of the fluxonium spectrum to $E_{J}$ and $E_{L}$
computed numerically for the $g\leftrightarrow e$ and $g\leftrightarrow f$
transitions and present device parameters.}
\label{FIG-Sensitivities}
\end{figure*}

\label{APP-T1}

\textit{Transitions }$e\rightarrow g$ $(g\rightarrow e)$\textit{.} Energy
relaxation of the qubit, taking into account the finite temperature of the
sample, an important effect around $m_{\mathrm{ext}}=1/2$, takes place at
the rate $\Gamma _{1}=\Gamma _{g\rightarrow e}+\Gamma _{e\rightarrow g}$,
where $\Gamma _{\alpha \rightarrow \beta }$ is the rate at which the qubit
makes a transition from states $\alpha $ to $\beta $. The relaxation time is
defined as $T_{1}^{eg}=1/\Gamma _{1}$ and is given in terms of the
black-sheep\ junction phase matrix element $\varphi _{eg}$ (Fig.~\ref%
{FIG-MatrixElements}b) and effective frequency-dependent parallel resistance 
$R_{\mathrm{eff}}(\omega )$ shunting the black-sheep\ junction. We may
decompose this resistance into the components $R_{\mathrm{intrinsic}}(\omega
)$ and $R_{\mathrm{Purcell}}(\omega )$, originating from the spontaneous
emission into the dissipative bath associated with the qubit circuit, and
into the measurement apparatus, respectively: 
\begin{equation}
R_{\mathrm{eff}}^{-1}=R_{\mathrm{intrinsic}}^{-1}+R_{\mathrm{Purcell}}^{-1}~.
\label{Reff}
\end{equation}%
The latter dissipation mechanism comes from the resonator-filtered $50$~$%
\Omega $ environment of the measurement circuit (Fig.~\ref%
{FIG-MeasurementSetup}a-b) and is called Purcell effect . The resulting
formula for $T_{1}^{eg}$ is obtained from Fermi's golden rule: 
\begin{equation}
1/T_{1}^{eg}=4\pi \frac{R_{Q}}{R_{\mathrm{eff}}(2\pi \nu _{eg})}|\varphi
_{eg}|^{2}\nu _{eg}\coth \frac{h\nu _{eg}}{2k_{B}T}~.  \label{T1Theory}
\end{equation}%
Here, 
\begin{equation}
R_{\mathrm{Purcell}}^{-1}(\omega )=Z_{\infty }^{-1}\frac{(\omega Z_{\infty
}C_{c})^{2}}{Q_{\mathrm{ext}}}\frac{\pi }{4}\frac{\left( \omega /\omega
_{0}\right) ^{2}}{\cot ^{2}(\frac{\pi }{2}\omega /\omega _{0})}~,
\label{T1Purcell}
\end{equation}%
%
%
%
%
%
%
%
%
%
%
%
%
%
%
%
%
%
%
%
%
%
is the real part of the admittance of the electrical circuit connected to
the black-sheep junction. In\textbf{\ }Fig.~\ref{cQEDschematic}a this is a
circuit connected to the two terminals of the element $Z^{\left\vert \alpha
\right\rangle }(\omega )$ with a pair of coupling capacitances $2C_{c}~$(the
capacitances $2C_{c}$ are also shown as the interdigitated capacitances in
Fig.~\ref{Figure1}a and Fig.~\ref{FIG-DevicePhotograph}c). The above
expressions work as long as $|1-\nu /\nu _{0}|\gg Q_{\mathrm{ext}}^{-1}$.
Note that $R_{\mathrm{Purcell}}\sim 1/C_{c}^{2}$, and that the value of $%
C_{c}$ in our circuit is of order $1~\mathrm{fF}$\textrm{, }making the
Purcell contribution smaller than in transmon qubits ($C_{c}\approx 20~%
\mathrm{fF}$) by more than two orders of magnitude. For the parameters of
our sample, the Purcell contribution becomes negligible as soon as $\nu
_{0}-\nu _{eg}>300~\mathrm{MHz}$. Therefore, energy relaxation in our qubit
is mostly intrinsic. Using the reasoning in the main text, we express $R_{%
\mathrm{intrinsic}}$ through the effective (and also frequency-dependent)
shunting resistances $R_{j}$ of the junctions (Fig.~\ref{Figure1}b):

\begin{equation}
R_{\mathrm{intrinsic}}^{-1}=R_{0}^{-1}+\left(
\sum\limits_{j=1}^{N}R_{j}\right) ^{-1}~.  \label{Rintrinsic}
\end{equation}%
Since the area of the black-sheep\ junction is only a factor of $6-7$
smaller than that of the array junctions, and assuming that $R_{j}^{-1}$ is
proportional to the area, it is likely that for large $N$, only the
black-sheep\ junction ($j=0$) contributes, $R_{\mathrm{intrinsic}}\approx
R_{0}$.

\textit{Relaxation }$f\rightarrow g$. In the vicinity of $m_{\mathrm{ext}}=0$%
, we have $\Gamma _{e\rightarrow g}\gg \Gamma _{f\rightarrow e}$, $\Gamma
_{f\rightarrow g}$. In the two-step transition $f\rightarrow e\rightarrow g$
the step, $f\rightarrow e$, is the bottleneck (Fig.~\ref{FIG-MatrixElements}%
b, \ref{FIG-ThreeLevelRelaxation}). Therefore, the relaxation time
associated with the $f\rightarrow g$ transition can be written as $%
1/T_{1}^{fg}\simeq \Gamma _{f\rightarrow e}+\Gamma _{f\rightarrow g}$, and,
neglecting the Purcell contribution, we get

\begin{align}
\left( T_{1}^{^{fg}}\right) ^{-1}& \simeq 4\pi \frac{R_{Q}}{R_{\mathrm{%
intrinsic}}(2\pi \nu _{fg})}|\varphi _{fg}|^{2}\nu _{fg}+
\label{T1threelevel} \\
& +4\pi \frac{R_{Q}}{R_{\mathrm{intrinsic}}(2\pi \nu _{fe})}|\varphi
_{fe}|^{2}\nu _{fe}\frac{1}{1-\exp \left( -\frac{h\nu _{fe}}{k_{B}T}\right) }%
~.  \notag
\end{align}%
We dropped the temperature-dependent factor in the first term because, in
our experiment, in the vicinity of $m_{\mathrm{ext}}=0$, the transition
energy are such that $h\nu _{fe}\approx k_{B}T\ll h\nu _{fg}$. Once $%
k_{B}T\simeq h\nu _{fg}$, the relaxation process becomes more complicated.

\begin{figure*}[tbp]
\centering
\resizebox{0.8\linewidth}{!}{\includegraphics{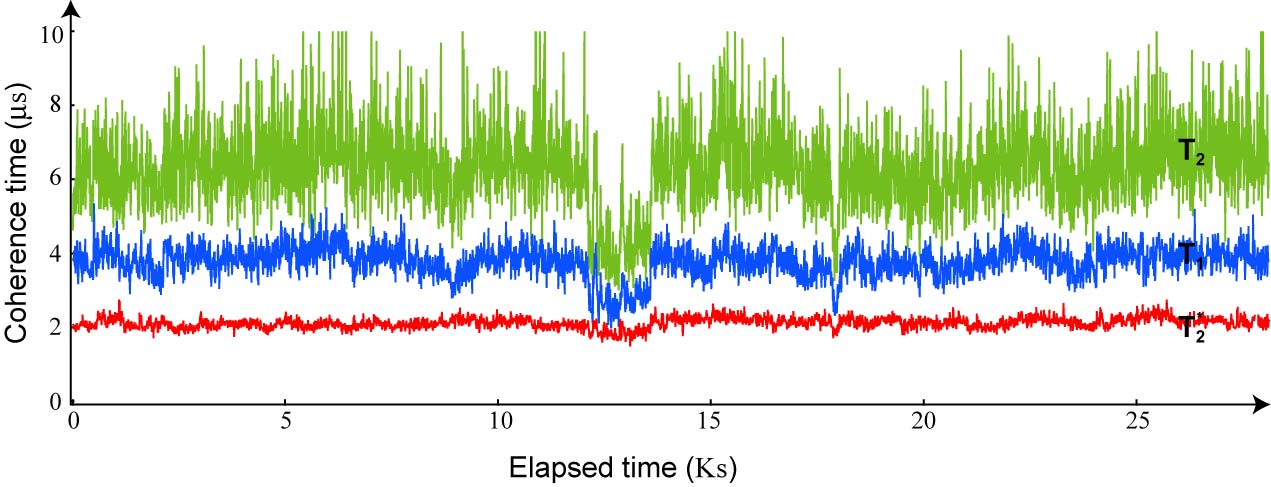}}
\caption{Measurement of $T_{1}$ (blue trace), $T_{2}^{\ast }$ (red trace)
and $T_{2}$ (green trace) as a function of time for the $g\leftrightarrow e$
transition at $|m_{\mathrm{ext}}|=0.2$. A set of the three data points was
acquired continuously every $8$ seconds.}
\label{Fig-TimesStability}
\end{figure*}

\subsection{Common dephasing mechanisms}

\label{APP-CDM}

\textit{Noise in }$m_{\mathrm{ext}}$. We establish a higher bound on its
amplitude, assuming the noise is \textquotedblleft 1/f\textquotedblright\ ($%
S_{m_{\mathrm{ext}}}=\delta m_{\mathrm{ext}}^{2}/\nu $, where $\nu $ is the
frequency in $\mathrm{Hz}$), from pure dephasing time at $m_{\mathrm{ext}%
}=0.2$, where the $T_{2}^{\ast }$ is maximal (Fig. 3a) but Aharonov-Casher
contribution is suppressed. For a \textquotedblleft 1/f\textquotedblright\
flux noise the pure dephasing time (gaussian decay of echo signal) is given
by~\cite{Nakamura} 
\begin{equation}
1/T_{\phi }^{m_{\mathrm{ext}}}\approx \delta m_{\mathrm{ext}}\times \mathrm{d%
}\nu _{eg}(m_{\mathrm{ext}})/\mathrm{d}m_{\mathrm{ext}}~.  \label{FluxNoise}
\end{equation}%
We estimate the dephasing time by subtracting the decay constant $1/T_{2}$
of the echo signal (which was nearly exponential), from the separately
measured $1/2T_{1}$: $T_{\phi }^{m_{\mathrm{ext}%
}}=(1/T_{2}-1/2T_{1})^{-1}>35~\mathrm{\mu s}$\textrm{.} Given that $\mathrm{d%
}\nu _{eg}(m_{\mathrm{ext}})/\mathrm{d}m_{\mathrm{ext}}\simeq 19.2~\mathrm{%
GHz}$ (Fig.~\ref{Figure2}b-c) we thus extract the \textquotedblleft
1/f\textquotedblright\ flux noise amplitude to be $\delta m_{\mathrm{ext}%
}<2\times 10^{-6}$.

\textit{Noise in }$E_{J}\equiv E_{J_{0}}$\textit{. }According to a previous
study\cite{vHarlingen}, this noise is believed to be \textquotedblleft
1/f\textquotedblright\ with the spectral density $S_{E_{J}}=\delta
E_{J}^{2}/\nu $, where $\delta E_{J}\propto E_{J}$. To first order, the
dephasing time $T_{2}$ due to fluctuating $E_{J}$ is proportional to $\left(
h\mathrm{d}\nu _{eg}(m_{\mathrm{ext}})/\mathrm{d}E_{J}\right) ^{-1}$.
Remarkably, present fluxonium circuit is supposed to be insensitive to the
noise in $E_{J}$ for some specific value of $m_{\mathrm{ext}}\simeq 0.43$
(Fig.~\ref{FIG-Sensitivities}a). Overall theoretical non-monotonic
dependence of $T_{2}$ on flux makes the noise in $E_{J}$ completely
incompatible with the data. For instance, if the $E_{J}$ noise limits
dephasing time at $m_{\mathrm{ext}}=1/2$ to the measured value of $250~%
\mathrm{ns}$, then it should also limit it to a similar number at $m_{%
\mathrm{ext}}\simeq 1/4$, where we measure the dephasing time one order of
magnitude longer. By analogy with the flux noise relation (\ref{FluxNoise}),
we can estimate $\delta E_{J}$ from the similar relation 
\begin{equation}
1/T_{\phi }^{E_{J}}=\delta E_{J}\times \mathrm{d}\nu _{eg}(m_{\mathrm{ext}})/%
\mathrm{d}E_{J}~.  \label{EJnoise}
\end{equation}%
Assuming that at $m_{\mathrm{ext}}=0.2$ the pure dephasing is entirely due
to \textquotedblleft 1/f\textquotedblright\ $E_{J}$ noise, we substitute $%
T_{\phi }^{E_{J}}>35~\mathrm{\mu s}$ and $h\mathrm{d}\nu _{eg}(m_{\mathrm{ext%
}})/\mathrm{d}E_{J}\simeq 0.1$ (Fig.~\ref{FIG-Sensitivities}a), we extract a
conservative estimate $\delta E_{J}<3\times 10^{-5}E_{J}$.

\textit{Noise in }$E_{L}$. Fluctuations in the Josephson energies $%
E_{J_{j\neq 0}}$ of the array junctions would result in a noisy inductance
(noisy $E_{L}$) such that $S_{E_{L}}=\delta E_{L}^{2}/\nu $. Already from
the considerations of model A it is clear that $\mathrm{d}\nu _{ge}(m_{%
\mathrm{ext}})/\mathrm{d}E_{L}$ is minimal at $m_{\mathrm{ext}}=1/2$ and
maximal at $m_{\mathrm{ext}}=0$\textbf{, }see Eq.~(\ref%
{PhaseSlipsLinewidthSimple}) and Fig.~\ref{Figure1}f. Calculating this
derivative numerically (Fig.~\ref{FIG-Sensitivities}b) we conclude that the
theoretical flux-dependence of the $E_{L}$ noise indeed is completely
inconsistent with our dephasing data. We extract the higher bound on the $%
E_{L}$ noise, assuming it is 1/f, from the pure dephasing time of the $%
g\leftrightarrow f$ transition measured at $m_{\mathrm{ext}}=0$, where flux, 
$E_{J}$, and Aharonov-Casher dephasing effects are minimal (Fig.~\ref%
{Figure2}c, Fig.~\ref{FIG-ACdephasingTheory}, and Fig.~\ref%
{FIG-Sensitivities}a). Using, as usual, $1/T_{\phi
}^{E_{L}}=1/T_{2}-1/2T_{1}^{fg}$, the relation 
\begin{equation}
1/T_{\phi }^{E_{L}}=\delta E_{L}\times \mathrm{d}\nu _{fg}(m_{\mathrm{ext}})/%
\mathrm{d}E_{L}~,  \label{ELnoise}
\end{equation}%
the observed values $T_{\phi }>50~\mathrm{\mu s}$, and $h\mathrm{d}\nu
_{fg}(m_{\mathrm{ext}})/\mathrm{d}E_{L}\simeq 16$ (Fig.~\ref%
{FIG-Sensitivities}b) at $m_{\mathrm{ext}}=0$, we find $\delta E_{L}<3\times
10^{-6}E_{L}$.

\subsection{Time-stability of the decoherence times}

\label{APP-timesstability}

Fig.~\ref{Fig-TimesStability} shows that the measured values of $T_{1}$ may
show fluctuations by as much as $50\%$ over time. The data is taken for the
lowest $g\leftrightarrow e$ transition for $m_{\mathrm{ext}}=0.2$ and is
typical for any other bias. The origin of these $T_{1}$ fluctuations is
unknown. Fortunately, the dephasing times $T_{2}^{\ast }$, measured in a
Ramsey fringe experiment, being significantly lower than $2T_{1}$, are
almost unaffected by the fluctuations in $T_{1}$. We also note that the
times $T_{2}$, measured in an echo experiment, follow the fluctuations of $%
T_{1}$, confirming that the value of $T_{2}$ is indeed close to $2T_{1}$.

\end{document}